\documentclass[preprint,12pt,sort&compress]{elsarticle}
\usepackage{algorithm} 
\usepackage{algpseudocode} 
\usepackage{graphicx}
\usepackage{lineno}
\usepackage{amsmath}
\usepackage{booktabs}
\usepackage{adjustbox}
\usepackage{epstopdf}
\usepackage[utf8]{inputenc}
\usepackage[english]{babel}
\usepackage{textgreek}
\usepackage{textcomp}
\usepackage{hyperref}
\usepackage{color}
\usepackage{comment}
\usepackage{url}
\usepackage{soul}
\usepackage[table,xcdraw]{xcolor}
\usepackage{booktabs}
\usepackage[nottoc,numbib]{tocbibind}
\usepackage[colorinlistoftodos]{todonotes}
\usepackage{multirow}
\usepackage{changepage}
\hypersetup{
    colorlinks=true,
    linkcolor=blue,
    filecolor=magenta,      
    urlcolor=blue,
}
\usepackage[margin=1.25in]{geometry}
\journal{}

\begin{document}
\begin{frontmatter}

\title{Incorporating non-linear effects in fast semi-analytical thermal modelling of powder bed fusion}

\author{S. R. Cooke, C. W. Sinclair, D. M. Maijer}

\address{Department of Materials Engineering, The University of British Columbia, Canada}

\begin{abstract}

The usefulness of semi-analytical thermal models for predicting the connection between process, microstructure and properties in powder bed fusion has been well illustrated in recent years. Such an approach provides the promise of accuracy comparable to tools that are orders of magnitude more computationally expensive. The opportunity to make predictions that span several orders of magnitude in space and time comes at the cost of significant simplifications, limiting fully quantitative predictions without empirical calibration. This approach relies on solving a linear problem meaning that first order non-linear effects induced by e.g. the temperature dependence of material properties and surface boundary conditions, are not incorporated. Here, we revisit these limitations and highlight ways that temperature varying material properties and radiative heat loss from the melt pool can be systematically accounted for. These corrections, made with an eye to minimizing additional computational overhead, bring the technique’s predictive capability much closer to that of high fidelity thermal simulations. Quantitative comparisons to experiments are used to illustrate the important impact of including such corrections.

{\centering
\includegraphics[width=0.9\textwidth]{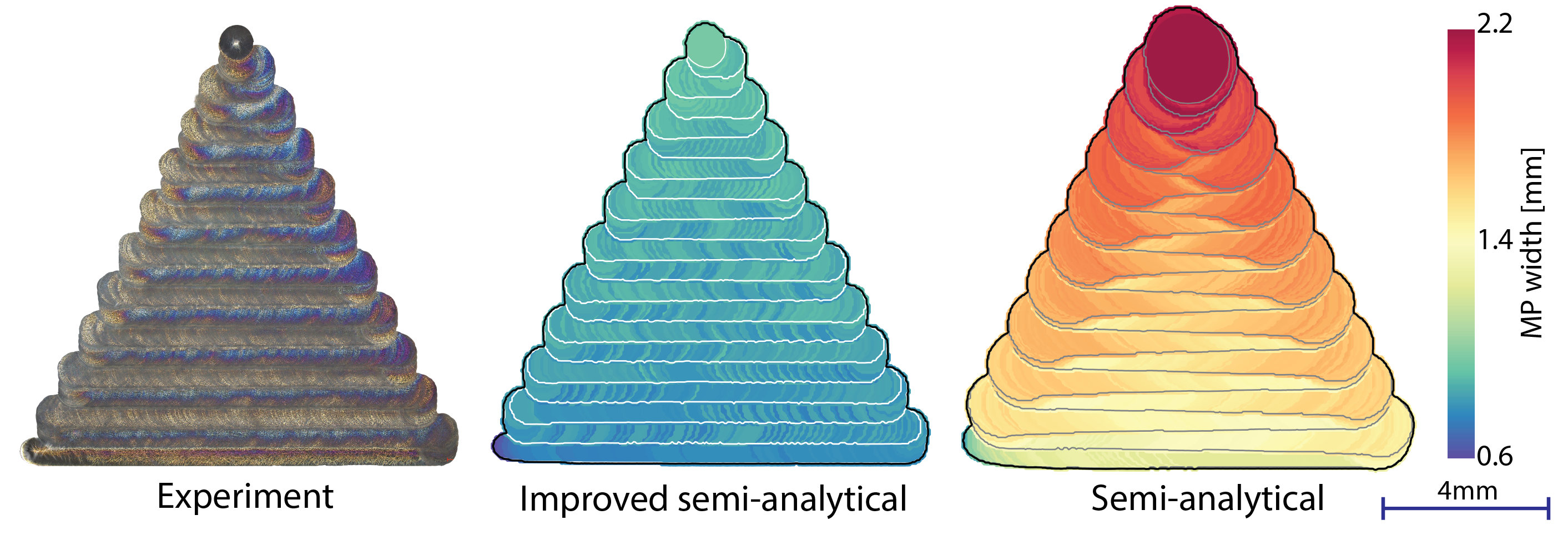}
\par
}

\end{abstract}


\end{frontmatter}

\section{Introduction}
\label{S:1}

Optimizing metal additive manufacturing (AM) faces combinatorial complexity, this arising from the vast process parameter space one can choose from. These choices have a significant impact on the quality of produced parts and so resource-intensive trial-and-error experiments are often conducted to help identify `safe' conditions (see e.g. \cite{bandyopadhyay2018invited}). Alternatively, one can avoid costly trials through the use of process models that attempt to reproduce key features of the process through \emph{in-silico} `experiments'. A hierarchy of numerical tools have been used in this regard; from high resolution computational fluid dynamics (CFD) simulations, solving mass, energy and momentum conservation  \cite{fotovvati2022multi,li2023particle,parivendhan2023numerical} through to low resolution thermal-only models, where significant coarse graining is used to allow fast predictions on large length and timescales \cite{stump2021solidification}. 

Fundamentally, the challenge is to be predictive over multiple orders of magnitude in length (µm - cm) and time (µs - hours). CFD provides high fidelity predictions but at a computational cost that limits one to length and timescales far from those needed to predict a full part. Given the history dependence of the predictions of such models, the quantitative translation of such predictions to a real part is a challenge \cite{plotkowski2023assessment}.  At the other end of the spectrum, the `super-layer' technique used in some thermal-only finite element (FE) simulations provides an opportunity to approximate the long range thermal field at part scale. This level of coarse graining eliminates predictions of melt pool geometry or local temperature gradients and raises questions about the tradeoff between speed and accuracy \cite{singh2022thermo,peng2018fast,zhang2022finite}. Thermal-only FE models that sit at intermediate length and time scales attempt to resolve (in an approximate sense) melt pools but with simulations run over the scale of single layers or, potentially small, geometrically simple, parts \cite{plotkowski2023assessment,schoinochoritis2017simulation}.  Such simulations have become the best option for many process predictions, but these remain fundamentally limited in their usefulness as the computational cost remains too large for most practical real part-scale simulations. Ultimately, one would like to extend these thermal-only simulations to be more computationally efficient, without further reducing their predictive capabilities.

One approach to the challenge of reducing the computational cost of such thermal simulations is the construction of purely data driven models\cite{bayat2021review}. These approaches (see e.g. Ref. \cite{lee2019blockchain}) can provide speedup but at the cost of large training sets and the associated cost of producing them. Alternatively, surrogate or reduced order models can be constructed (e.g. \cite{li2022time}) which attempt to either speed up the process of numerically solving the partial differential equations in the FE model (see e.g. \cite{li2022time,zhao2021enhancing})  or by using regression models on reduced order descriptions of training data (see. e.g.\cite{li2022time}). In these cases one can achieve speed up but typically at the cost of accuracy (in the first case) or with the need for large training sets (in the second case).     

An alternative, older, approach that has seen renewed interest is the semi-analytical approaches long known from the welding literature (e.g. \cite{eagar1983temperature}). Starting from the assumption of temperature independent material properties, semi-infinite solution domain and a point heat source, one obtains the Greens Function solution to the heat equation. As this is the solution to a linear problem, combinations of this solution are themselves a solution, meaning that one can quickly and easily build up solutions for moving heat sources of arbitrary shape by replacing point heat sources with sources distributed as functions in space and time. The classic Rosenthal solution is a limiting case where the heat source is assumed to be a single point moving at constant velocity in the limit of infinite time (steady-state) \cite{rosenthal1941mathematical,rosenthal1946theory}.  

The result of such a semi-analytical model is an equation providing an explicit relationship between the temperature at a location in space as a function of time. In general, the result remains semi-analytical because an integral over time must be performed, which in the general case, does not have an analytical solution. Without the need to solve the partial differential equation numerically, one is no longer required to approximate gradients, this removes mesh sensitivity or time step limitations. This opens up the possibility, revealed through the work of Plotkowski \emph{et al.} \cite{plotkowski2019influence,plotkowski2021stochastic,plotkowski2017verification}, to have a tool that can make predictions of microstructure as a function of arbitrarily complex beam trajectories within complex shaped parts. 

The assumptions underlying this semi-analytical approach mean that it is limited in its quantitative predictions compared to higher computational cost techniques \cite{stump2021solidification}.  Some of the above limitations can be addressed approximately.  For example, adiabatic boundary conditions can be easily incorporated using the method of images \cite{steuben2019enriched}, this again taking advantage of the additivity of solutions. The more difficult limiting assumptions, and the ones that most fundamentally limit the predictive power of this technique compared to numerical solutions of the heat equation, are the non-linearities associated with heat loss from the melt pool and the temperature dependence of the materials properties. Attempts have been made for specific materials and specific build conditions to identify the best matching constant material properties for fitting experimental \textcolor{blue}{results \cite{yang2018semi,plotkowski2017verification,nandwana2020predicting}.} In some cases this was shown to result in unrealistic values of the thermal conductivity likely owing to the fact that the optimization was conducted without considering the effects of heat loss from the melt pool in parallel \cite{yang2018semi}. To avoid the need for calibration, Steuben \emph{et al} \cite{steuben2019enriched} implemented a recursive model to approximate the temperature dependence of the material properties. While the results appeared promising, they too were obtained without considering heat loss from the melt pool and results were presented for only a limited number of cases.

To address surface heat losses, Shahabad \emph{et al} \cite{imani2021extended} adopted an approach of empirically scaling the input power to account for losses due to convection and radiation, radiation being dominant for all cases studied. However, their model assumed steady-state conditions and tested only a limited range of processing parameters. Li \emph{et al} \cite{li2023semi} showed similar effects using a semi-analytical model derived through separation of variables while  Ning \emph{et al} \cite{ning2019analytical,ning2019analyticalti64,ning2019analytical1,ning2019analytical2} accounted for heat loss through stationary heat sinks. In this case, while the results for melt pool geometry compared favourably with those from the FE method, the long range temperature field was poorly reproduced. In all cases, the problem of heat loss was tackled without dealing with the parallel problem of the temperature dependence of the material properties. 

In this work, we revisit the semi-analytical thermal model developed by  Tsai and Eager \cite{eagar1983temperature} and build on the recent work of Stump et al \emph{et al} , with the aim  of systematically and simultaneously correcting for the non-linear effects arising from  surface heat loss and temperature dependent material properties. Our aim here is  to show that one can systematically improve the semi-analytical solution in a way  that can be generalized to any material used in metal powder bed fusion AM without  the need for empirical corrections. Our goal is to achieve quantitative predictions  with good agreement compared to those obtained from much more computationally  expensive finite element simulations. In this regard, we systematically compare our \emph{Fast-running semi-Analytical Self-consistent Thermal} (FAST) model to FE simulations for a range of transient conditions. We also show a comparison against simple  laser remelting experiments and discuss the ways in which this approach allows for  massive speedup at low cost in accuracy.

\section{Implementation of Numerical Models}

As noted above, the aim of this work is to correct the classical semi-analytical solution so as to reproduce predictions obtained using a full FE solution calculation, including the non-linear effects that are absent in the classic semi-analytical approach. In this regard, we present two types of numerical predictions in this work, those coming from our implementation of the (corrected) semi-analytical formulation and those coming from thermal-only FE calculations.  The details of how these two sets of calculations were performed are presented in this section.

While our target is to simulate powder bed additive manufacturing, we make certain simplifications to focus on a first order implementation. Thus, the work focuses on predictions related to remelting of a solid surface. However, as has been discussed recently \cite{upadhyay2023effect}, so long as the depth of the melt track is larger than the powder layer thickness over a solid part then the inclusion of powder properties has little to no effect on the prediction of melt pool geometry and/or long range temperature fields from FE simulations. We also do not discuss the implementation to multilayer builds as this will be left to a future discussion. Finally, we have previously discussed the implementation of latent heat into the semi-analytical formulation as an additional Gaussian heat source \cite{cooke2023enhancement}. To retain focus on the non-linear effects of surface heat losses and temperature dependent material properties, we do not include latent heat into our calculations.

In all cases, the simulations were performed assuming an isotropic, symmetric Gaussian heat source that imposes a surface flux on the material.  The beam width (standard deviation) was assumed to be $\sigma_{xy} = 0.145$~mm in all simulations, this being consistent with the values used in L-PBF and EBAM simulations \cite{carriere2018energy}. An absorption efficiency of $\eta = 0.72$ has been assumed, this being consistent with simulations of experiments under electron beam powder bed processing \cite{carriere2018energy}. Since this is assumed to scale the input beam power $Q$, it does not fundamentally affect the conclusions presented here.

\subsection{Implementation of the Semi-Analytical Solution for a Moving Gaussian Heat Source}

The starting point for this work is the semi-analytical model originally proposed by Tsai and Eager \cite{eagar1983temperature,woodard1999thermal, nguyen1999analytical} (Equation~\ref{eqn:nguyen}).
\begin{equation}
    T(t)-T_{0}=\frac{2 \eta}{\rho c_{p}(\pi / 3)^{3 / 2}} \int_{0}^{t} \frac{Q(t')}{\sqrt{\phi_{x} \phi_{y} \phi_{z}}} \exp \left(\frac{3 x \left(t^{\prime}\right)^{2}}{\phi_{x}}-\frac{3 y \left(t^{\prime}\right)^{2}}{\phi_{y}}-\frac{3 z\left(t^{\prime}\right)^{2}}{\phi_{z}}\right) d t^{\prime}
    \label{eqn:nguyen}
\end{equation}

This expression describes the temperature, $T$, as a function of the time, $t$, at a point of interest ${x_p,y_p,z_p}$ in space, given a Gaussian shaped heat source with a power, $Q$, a beam center located at ${x_b(t'), y_b(t')}$, and moving at a velocity, $v_B$. The position $x(t')$ is then defined as $x_p$ - $x_b(t')$, where $t'$ is the variable of integration. Similarly, the same applies for $y(t')$ and $z(t')$. Because the integral depends explicitly on the time at which we want the solution at, the integral must be performed explicitly from $t' =0$ to $t' = t$ for each distinct value of $t$. 

The solution requires that the thermophysical properties, namely the specific heat, $c_p$, thermal conductivity, $k$, and, density, $\rho$, are constant and independent of temperature, and that heat transfer occurs by conduction-only within a semi-infinite domain. It is assumed that the temperature throughout the domain is uniform and equal to $T_{0}$ at time ($t'$) = 0~s. The parameters $\phi_{i}$ = 12$\alpha (t-t') + \sigma_{i}^{2}$ for $i=x,y,z$, include the (assumed) constant thermal diffusivity $\tilde{\alpha}$ and the standard deviation of the Gaussian heat source ($\sigma$). If $\sigma_x$ and $\sigma_y$ are set independently then one produces an elliptical beam cross-section while setting $\sigma_x = \sigma_y = \sigma_{xy}$ for a beam diameter of 0.145 mm, as done here, produces a symmetric Gaussian beam shape. Assigning a value to $\sigma_z$ allows for heat to be deposited within the thickness of the material whereas when $\sigma_z \rightarrow 0$, the volumetric heat source becomes a 2D surface heat source. In this work, $\sigma_z$ is set to zero which is consistent with the FE simulations.

Equation 1 can be imagined as a sum of solutions infinitesimally shifted in time and space. The solution for a moving Gaussian heat source (a Gaussian itself) is illustrated in Figure 1 as the result of a series of contributions from Gaussian heat sources based on where the beam has previously been located. Each of these decaying Gaussian solutions evolves independently with time. In the limit of an infinitesimal timestep between each independent source, the sum becomes an integral giving the exact temperature as a function of position at the current time $t$.
    
    \begin{figure}[!h]
    \centering
        \includegraphics[width=0.8\textwidth]{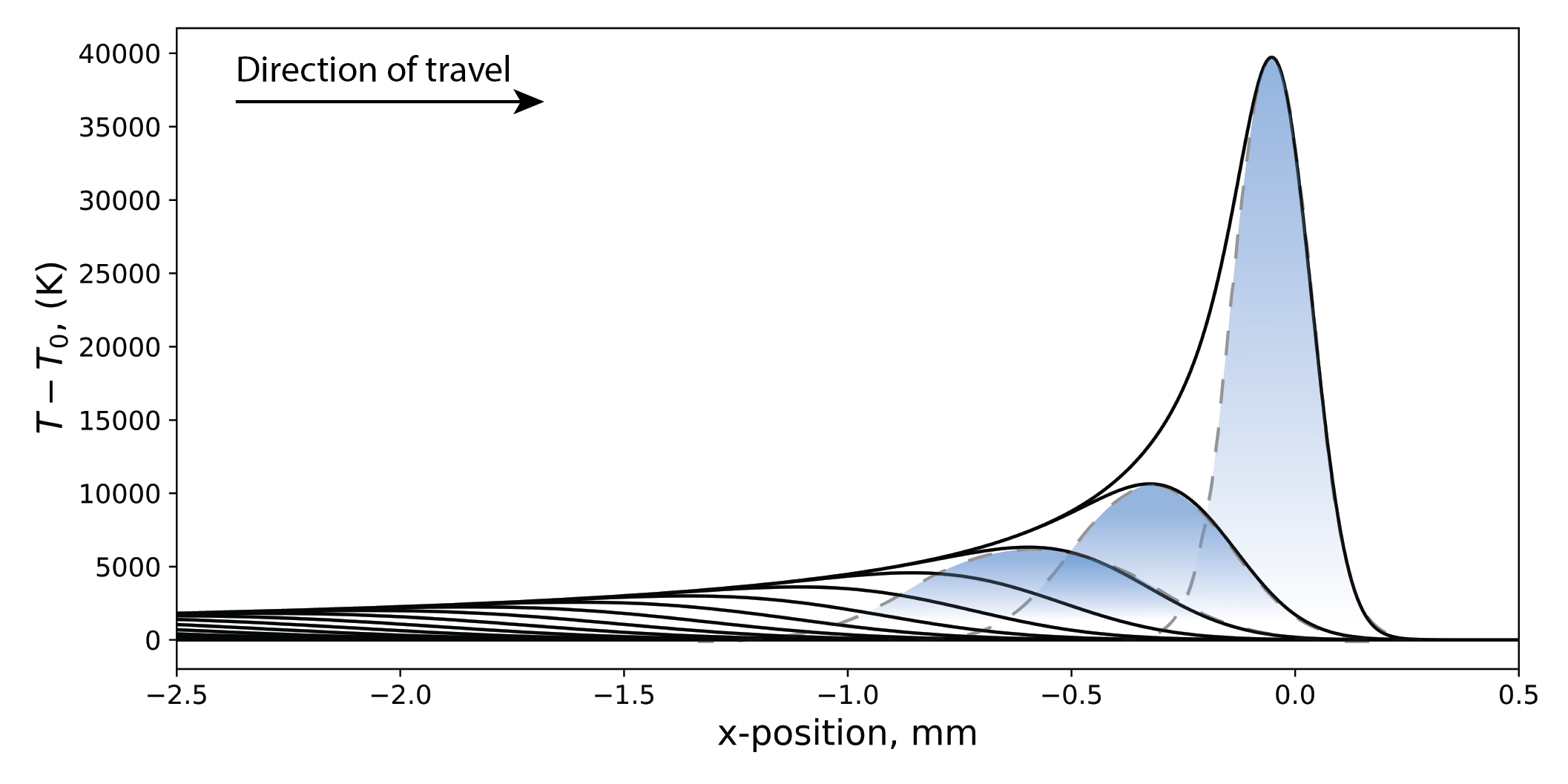}
        \caption{An illustration of equation \ref{eqn:nguyen} showing how it can be seen as the sum of a series of independently decaying Gaussians separated in time and space based on the beam motion. Here, the solution is shown for the steady-state profile at a power of 300W, a beam velocity of 50mm/s for Ti-6Al-4V.}
        \label{fig:SA_sums}
    \end{figure}  
    
The FAST model reads the user-defined beam positions, ${x_b,y_b,z_b}$, and power, $Q$, as basic G-code and performs the calculation of the temperature at user identified positions ${x_p,y_p,z_p}$ as a function of user selected times. The main bottleneck in the calculation is the integration of equation \ref{eqn:nguyen}.  Following  Stump \emph{et al} \cite{stump2019adaptive}, we have implemented a Gaussian-Legendre integration scheme to numerically compute  the integral. A constant integration segment length was determined based on the beam velocity to provide a smooth curve. CPU parallelization was implemented using OpenMP \cite{openmp} to compute the integral simultaneously for the spatial locations at each timestep.

\subsection{Finite Element Simulations:}

\begin{figure}[!h]
    \centering
        \includegraphics[width=1\textwidth]{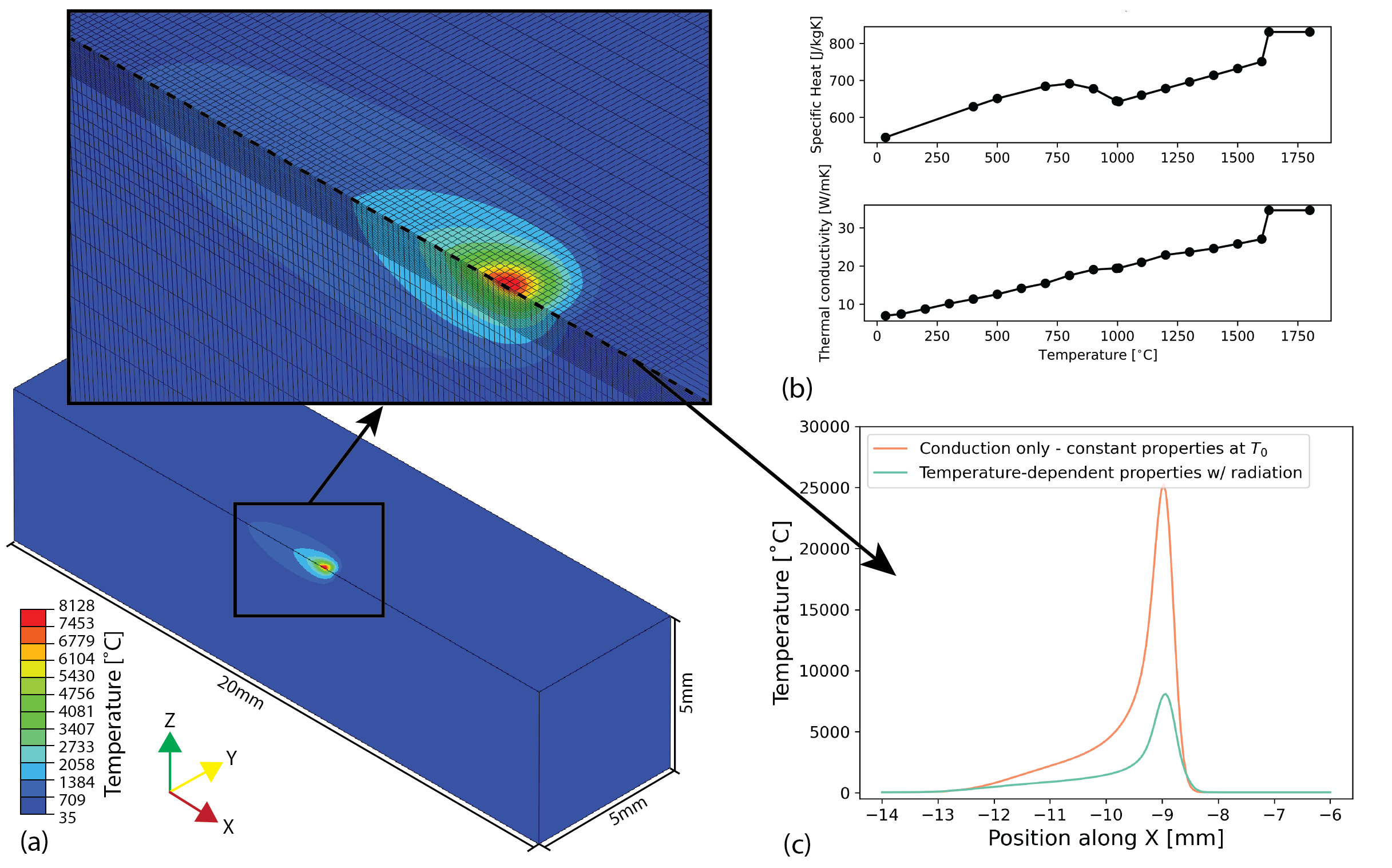}
        \caption{Finite element (FE) analysis details showing a) the full FE domain used in this study, highlighting the dimensions of the mesh close to the area of interest, b) the temperature-dependent material properties used for Ti-6Al-4V \cite{zhao2014three}, and c) temperature profile along the center-line of the melt track highlighting the differences in temperature prediction using i) constant material property values (evaluated $T_0$ = 35$^\circ$C) and no radiation boundary condition versus ii) simulations including both radiation and temperature-dependent material properties. Simulations performed for a power of 300W and a beam velocity of 50mm/s.}
        \label{fig:FE_compare}
    \end{figure}

The FE model used here was built in ABAQUS where  temperature-dependent heat capacity and thermal conductivity for Ti-6Al-4V were used \cite{zhao2014three}, see Figure~\ref{fig:FE_compare}. A constant density of 4200 $kg/m^3$ was used to avoid the need to solve the coupled thermal-mechanical equations and the creation of mass on a fixed domain volume. A moving Gaussian heat flux with a circular beam diameter was applied to the heated surface using the DFLUX user-subroutine in ABAQUS. A radiation boundary condition was also applied on the top surface of the domain, using an emissivity of 0.7 \cite{carriere2018energy}, where radiation is assumed to occur to the surrounding environment at a temperature of 30$^\circ $C. For the transient heat transfer analysis, 242,172  DC3D8 elements were constructed from 257,312 nodes. The finest mesh, covering the area around the affected domain heated by the beam, is comprised of elements having dimensions of 42 x 42 x 10 $\mu m$. The mesh size was systematically increased away from the center-line of the beam trajectory coinciding with the decreasing thermal gradient to reduce computation time. A mesh sensitivity study was conducted to ensure mesh size independence at the mesh size used here. To further reduce computation time in the FE model, a symmetry boundary condition (adiabatic surface) was assumed at the central plane of the domain. Lastly, the domain was made large enough to approximate semi-infinite conditions for the processing parameters tested. The FE domain, mesh and comparison between the temperature profile for both temperature-dependent properties with radiation and the assumptions without radiation heat transfer and constant properties evaluated at the initial temperature are shown in Figure~\ref{fig:FE_compare}.

\clearpage

\section{Correcting for Non-Linear Effects Arising from Material Properties and Radiative Losses}

\subsection{Temperature Dependence of Material Properties}

As noted in the introduction, one of the necessary but oversimplifications required in developing the semi-analytical solution presented in equation \ref{eqn:nguyen} is that the material properties (density, specific heat capacity and thermal conductivity in this case) are constant. As highlighted in Figure~\ref{fig:FE_compare}(b) for Ti-6Al-4V, and in general for most alloys, this is a poor approximation and one that introduces significant disagreement between full numerical solutions of the heat equation that include temperature dependent properties, and those that don't. Here we ask whether there is an optimal way to choose the material property values so as to best approximate the true FE temperature field. The results in Figure~\ref{fig:FE_compare}(c), for example, were obtained using constant property values evaluated at the initial background temperature, $T_0$ = 35$^\circ$~C. As this is far from the liquidus temperature ($T_L$), it is unsurprising that the predictions close to the melt-pool are poor, yet far from the melt pool (where $T\rightarrow T_0$), one finds that the temperature predictions are much better.  In contrast, if one evaluates the properties at $T_L$ then (for the case of Ti-6Al-4V), one would find that the melt pool size and far field temperatures are under-predicted owing to the fact that the thermal properties are over-estimated throughout the majority of the simulation domain.

The need for an intermediate approach has previously been recognized in the welding literature where material property averages calculated based on the temperature range experienced by the part have been shown to provide a much better prediction compared to using thermal properties evaluated at any one specific temperature \cite{zhu2002effects}.  This, requires, however, the full thermal history to be known \emph{a priori}. A much simpler approach is to use a temperature-averaged property value \cite{mendez2023calculation} which can be easily calculated prior to the start of the simulation. For example, the thermal diffusivity can be estimated in this way as,

\begin{equation}
    \bar{\alpha}=\frac{\int_{T_0}^{T_U} \alpha\left(T^{\prime}\right) d T^{\prime}}{\int_{T_0}^{T_U} d T^{\prime}}
\label{eqn:props}
\end{equation}

\begin{figure}[!h]
    \centering
        \includegraphics[width=1\textwidth]{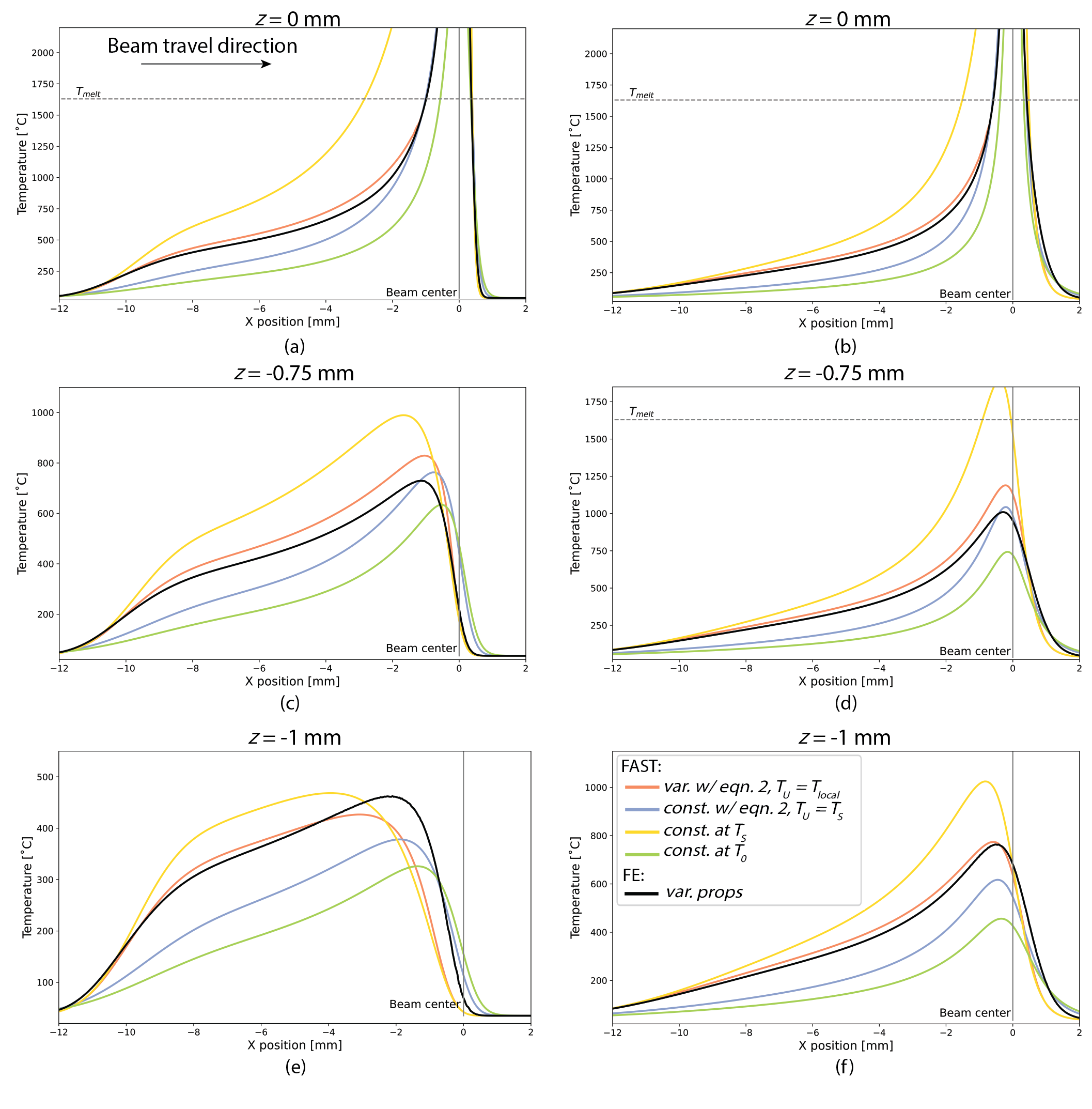}
        \caption{Conduction-only steady-state temperature profiles calculated with the FAST model for Ti-6Al-4V properties with an initial substrate temperature of $T_0$ = 35$^\circ$C, evaluated at $T_0$, $T_L$, and average properties calculated as $T_0 \rightarrow T_L$, and $T_0 \rightarrow T_{local}$, where $T_{local}$ is the node temperature at the previous timestep, based on Equation~\ref{eqn:props} compared to FE predicted temperature profile with temperature-dependent properties.  Process conditions of 300W at 50~mm/s are shown for a) along the surface, c) 0.75~mm below the surface and e) 1~mm below the surface. Similarly, for 150W at 8.3~mm/s, the results are shown for b) along the surface, d) 0.75~mm below the surface and f) 1~mm below the surface.}
        \label{fig:props_compare}
    \end{figure}

As $T_0$ increases towards $T_U$, the average value asymptotically approaches the value of the material properties evaluated at $T_U$. As a first approach in this work, $T_U$ is set to the liquidus temperature, $T_L$, in an attempt to capture the full temperature range of properties in the material expected to be sampled. For $T_0$ well below $T_L$, the average property is more heavily weighted towards lower temperatures. Figure \ref{fig:props_compare} compares the  temperatures predicted by the FAST model along lines centered on the melt track at the top surface, 0.75 mm below the top surface and 1 mm below the top surface. These are shown for steady state conditions at $Q$ = 300~W and $v$ = 50 mm/s (6~J/mm) and $Q$ = 150~W and $v$ = 8.3 mm/s (18~J/mm) with $T_0 = 35^\circ$C. The temperatures have been predicted for a variety of different assumptions for how the material properties have been evaluated. To observe just the effect of the material properties, radiative heat loss is neglected in all models for this subsection. As noted already, evaluating the properties at $T=T_0$ and $T=T_L$ provides significant over and under predictions of the temperatures, respectively. When $T=T_0$ is used to evaluate the properties, the temperatures close to the melt track are poorly predicted, while when $T=T_L$ is used, the long range temperature field is poorly predicted.  In contrast, when the properties are evaluated using the average in equation~\ref{eqn:props}, where $T_U$ = $T_L$, we see a much closer prediction across all values of $z$.

A more computationally expensive approach is to adjust the temperature weighted average (equation \ref{eqn:props}) to the range of temperatures expected in a given location.  Thus, in regions near the melt pool, the limits of integration would remain $T_0$ and $T_L$, while at positions far away from the melt track, one would limit the upper bound of the integration to $T_U < T_L$ where $T_U$ is the temperature at the point of interest from the previous timestep.

At first sight it might appear that this approach breaks the explicit assumption of constant material properties because we propose to vary the properties spatially within the domain. Recall, however, that the solution represented by equation \ref{eqn:nguyen} gives us the temperature at a given location as a function of time under the assumption that the rest of the domain has uniform material properties. Conceptually there is nothing restricting us from calculating the temperature at another location under the assumption that the domain has uniform material properties that are different from the first location.

Thus, if one wishes to track the temperature at a fixed location as a function of time, one can use the temperature calculated at that location in the previous timestep as the upper limit in the average (equation~\ref{eqn:props}). This requires an additional, though inexpensive calculation on each timestep. This method (shown as the orange lines in figure~\ref{fig:props_compare}) provides the best match to the FE predicted temperatures. In this example, we slightly over-predict temperatures in most locations. The largest deviation when using this last technique comes when we look at temperatures well below the top surface (e.g. at a depth of 1 mm) and in front of the beam position. For the process conditions at 300W and 50mm/s, we see a noticeable lag in the thermal response caused by the high thermal gradients. This is less apparent under the conditions of 150W and 8.3mm/s, due to the lower power and beam velocity and consequently, the decrease in thermal gradients around the melt pool. The rapid change in temperature means that our use of the temperature in the previous timestep under-predicts the current temperature and therefore (for Ti-6Al-4V) leads to a lag in thermal diffusion. It is possible to improve the prediction in this region by iterating, using the temperature from the previous iteration to improve the location specific prediction of the material properties, but this would come at significant computational cost and so has not been implemented here.
 
\clearpage

\subsection{Heat Loss by Radiation}
Surface heat loss is not expressly accounted for in the semi-analytical model but heat loss from the surface has a considerable effect on predicted temperatures, particularly at high energy densities and initial temperatures \cite{yang2018semi}. The impact of surface heat loss becomes increasingly important as one moves from single melt tracks to complex beam trajectories on the scale of a part where heat can accumulate over time. By not tracking the heat loss due to radiation over time, the temperature will continue to build, which can lead to compounding errors for predicting the thermal history in the build.

\begin{figure}[!h]
    \centering
        \includegraphics[width=1\textwidth]{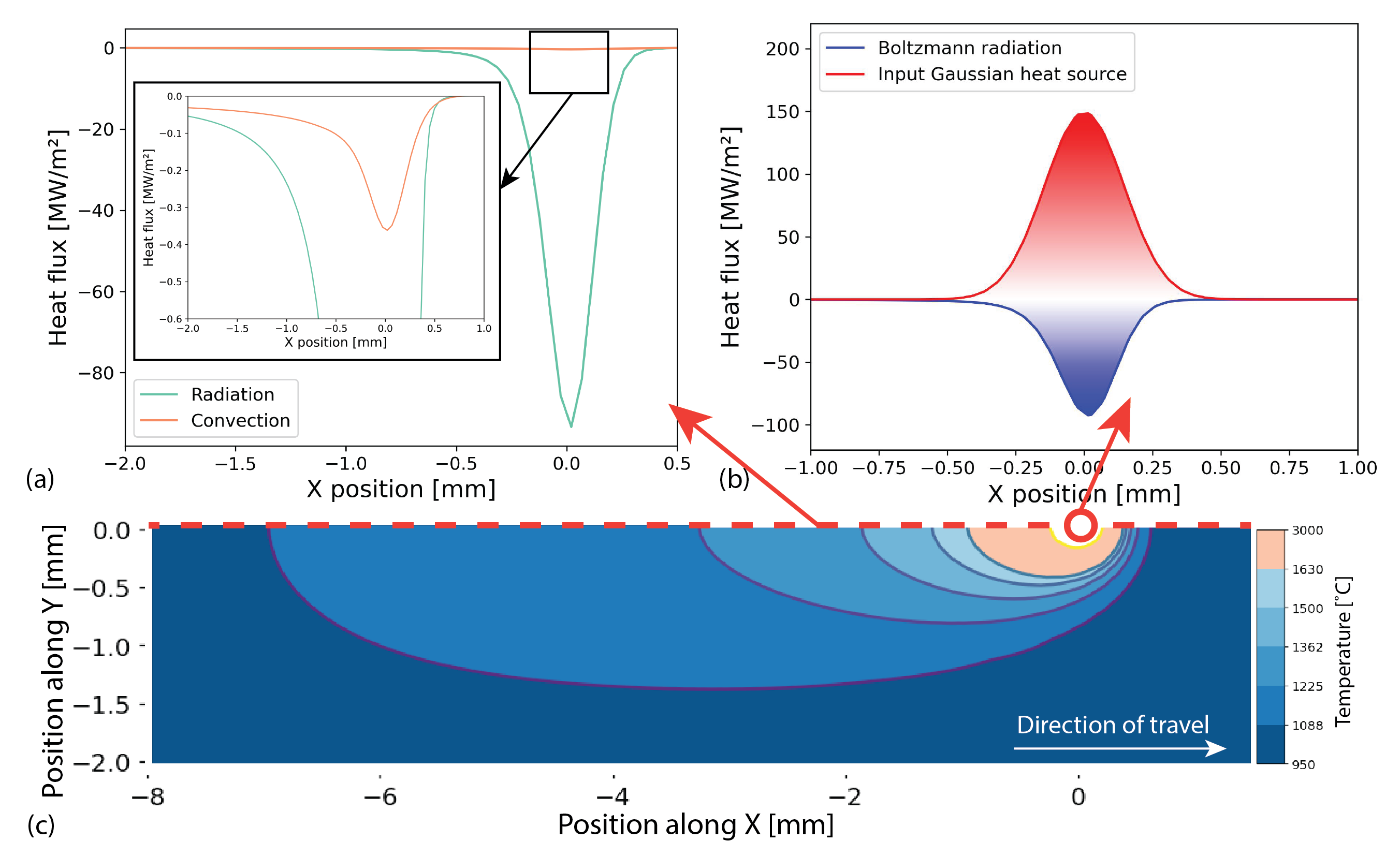}
        \caption{FE model with temperature dependent properties without radiation showing a) the difference in magnitude between the radiative and convective heat flux b) the comparison between the input Gaussian heat flux and the resulting Boltzmann radiative flux, and c) the surface temperature contour for a given set of processing parameters at steady-state with $Q$ = 150 W, $v_B$ = 50 mm/s and $T_0$ = 1000$^\circ$C}
        \label{fig:conv_compare}
    \end{figure}

Two forms of heat loss on the surface can be considered, depending on the conditions the process operates under; radiation or convection to the environment in the build chamber. In processes which have surface losses through convection (e.g. laser PBF), it is found that this is negligible compared to radiative losses for any reasonable value of the heat transfer coefficient. This is consistent with the findings of others (see e.g. \cite{imani2021extended,li2023semi}). Figure~\ref{fig:conv_compare}(a) illustrates this for a FE simulated single track, steady-state simulation where the radiative heat flux has been calculated using the Stefan-Boltzmann law and a forced convective film coefficient (100 $W/m^{2}K$) both based on the computed surface temperatures.

As noted above, others have found success in correcting for radiative losses in semi-analytical formulations through a simple empirical scaling of the beam power, $Q$ \cite{imani2021extended}. The fact that this should give good results is not, however, immediately obvious, nor is it clear whether this should be generally applicable. Figure~\ref{fig:conv_compare}(b) gives a partial explanation for why this may be an acceptable approximation under the conditions considered here.  In this figure we show the input power (red) and the heat loss from radiation calculated using the Stefan-Boltzmann Law based on the surface temperatures of the part. One sees that the radiative heat loss is approximately Gaussian in shape and approximately centred on the beam position. Summing the input and the loss is, therefore, nearly the same as simply scaling the input heat power by a value less than 1.   

\begin{figure}[!h]
    \centering
        \includegraphics[width=1\textwidth]{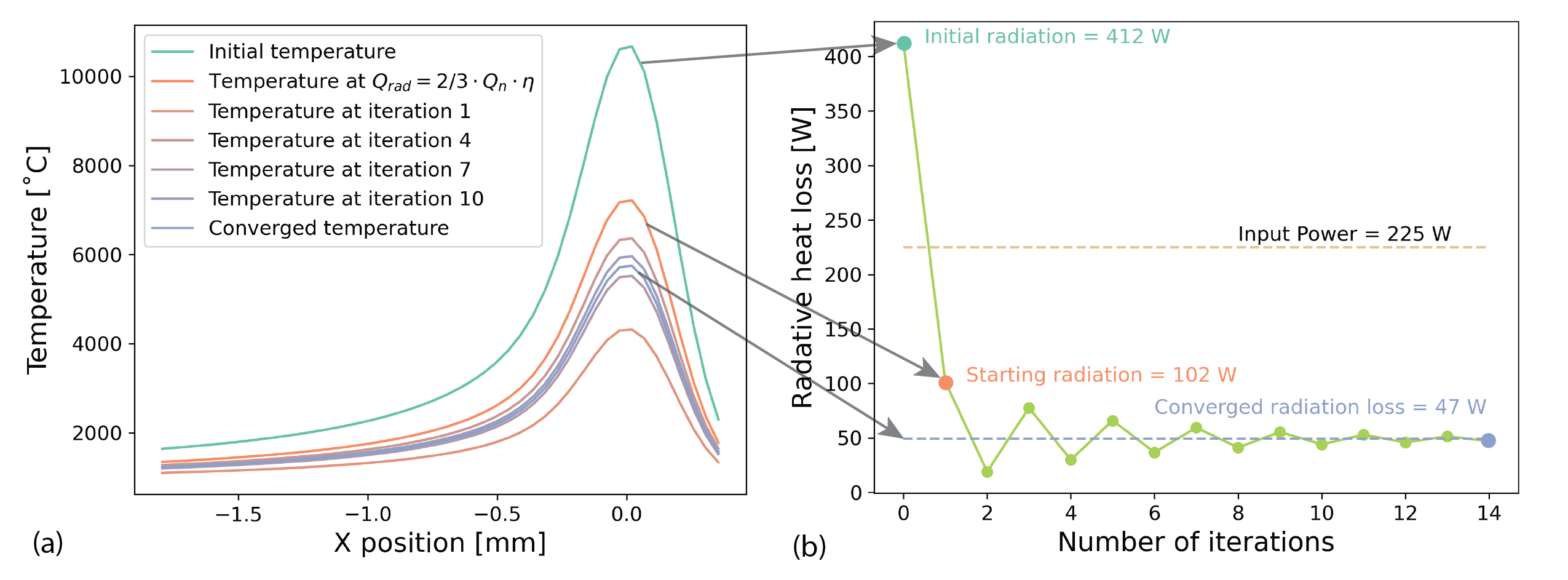}
        \caption{Results of the self-consistent iterative algorithm to approximate the radiation loss showing a) the 1D temperature profiles predicted by the FAST model at different stages of the convergence and b) the radiative heat loss corresponding to the temperature profiles. Process parameters for this example were calculated with $Q$ = 225W, $v_B$ = 50 mm/s and $T_0$ = 800$^\circ$C.}
        \label{fig:iter1}
    \end{figure}

Having established that one can approximately account for radiative losses by a simple scaling of the input power, the question of how to set its magnitude remains. Computing the temperature field without accounting for radiation and subsequently applying the Stefan-Boltzmann law to the resulting surface temperatures is inadequate as the temperature close to the beam center is significantly over predicted in most conditions \cite{mirkoohi2018thermal}. We have found, however, that the radiative losses can be effectively accounted for in a self-consistent manner by iteratively adjusting the radiative losses and the computed model temperature field (Figure \ref{fig:iter1}).  First, the temperature is computed neglecting radiative losses from the surface. From this temperature field, radiative losses are estimated using the Stefan-Boltzmann law  for all nodes on the melt pool surface. These are summed (weighted by the surface area associated with each node) to give the overall radiative heat loss. However, in doing so, one must ensure that enough nodes are used to accurately resolve the surface area of the meltpool in order to approximate the heat lost to radiation. As illustrated in Figure \ref{fig:iter1}(b), in this first iteration, the radiative losses may be larger than the input power.  If this is the case, the radiative loss is taken to be a fraction (here 2/3, though the exact value is not critical) of the input power, this being subtracted from the input power to give a new radiation-corrected input power. With this new input power, the temperature field is re-calculated as is the radiative loss from it.  This process is continued iteratively until the temperature field and corresponding radiative loss converges to a selected convergence limit. This iterative scenario is shown in Figure \ref{fig:iter1} for a steady-state temperature profile to illustrate this process.   

As previously discussed, we follow \cite{stump2019adaptive} in using Gaussian-Legendre integration to compute equation \ref{eqn:nguyen}. In performing the integration over time, the time history evolution of the radiative losses needs to be accounted for. In this case one might question the computational efficiency of a technique that requires several iterations for each timestep to achieve convergence.  However, so long as the beam's position and input power are smoothly varying this does not introduce noticeable computational cost as the radiative flux computed for the current timestep is used as the initial guess for the next. This avoids the need to repeat the integration after applying the radiation loss within the same time increment. A sensitivity analysis on the impact of timestep size was conducted verifying that as long as a smoothly varying temperature prediction was made, less than 5 iterative steps were required to converge the radiative losses. In the case where discontinuous melt pools are desired (see e.g. \cite{plotkowski2021stochastic}) then the cost would slightly increase as the iterative scheme would take more iterations to converge.

This iterative, self-consistent scheme for radiative heat loss has been tested across a wide range of process conditions and a variety of materials and have found it to be a very robust and computationally efficient method for systematically accounting for radiative losses.  As will be shown below, the power of this approach becomes more apparent when we apply the technique to highly transient conditions where the background temperature is increased over time due to the motion of the beam.

\section{Moving Beyond the Steady-State: Predictions Under Highly Transient Conditions}

\begin{figure}[!h]
    \centering
        \includegraphics[width=1\textwidth]{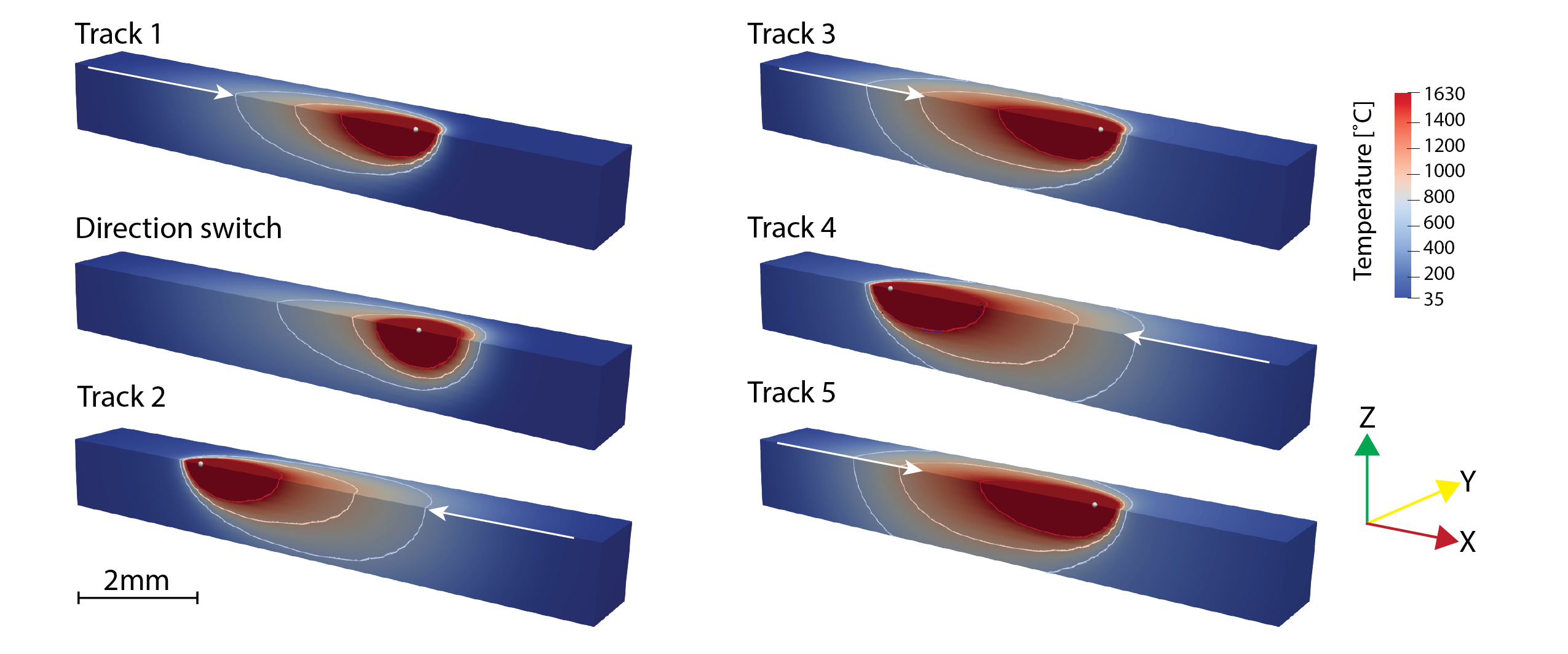}
        \caption{Temperature predictions for a case with back-and-forth beam motion computed with the FAST model used to evaluate different processing parameters and emulate the characteristics of cyclic heating and cooling within a small domain. The snapshots here shows a sub-section of the domain (the beam position is marked with a small white circle) at different beam positions and times. The figures labeled `Track 1',`Track 3' and `Track 5' all show temperatures near the end of the first, third and fifth pass (beam moving to the right) while figures labeled `Track 2' and `Track 4' show temperatures at the end of passes with the beam moving to the left. One can see, by comparing Tracks 1 and 5 how heat builds up with time close to the beam path.}
        \label{fig:path}
    \end{figure}

As previously discussed, the benefits of semi-analytical models are their ability to provide predictions over large length and timescales under highly transient conditions, rather than being computationally limited to single, short, melt tracks at steady state conditions. As a first step in this direction, while still allowing for a direct comparison with FE simulations, a simple back-and-forth beam motion for 5 passes along the same 4~mm long path is simulated.  The simulation domain is the same as the one shown in Figure~\ref{fig:FE_compare} but during the back-and-forth motion of the beam, heat accumulates close to the melt track (Figure~\ref{fig:path}). Thus, by repeatedly re-heating over the same track, this beam path accentuates challenging features relevant to AM, e.g. cyclic heating/cooling and the long range build up of heat over time. For these calculations, the radiative heat-loss correction described above and the material properties based on the temperature weighted average between $T_0$ and the local temperature at the previous timestep, $T_{local}$ have been implemented. It was found that, for the duration of the simulations performed (maximum $\sim$ 3~s), the temperatures at the boundary of the simulation domain did not change appreciably, this allowing us to use a semi-infinite sized domain in our FE simulations.  

\begin{figure}[!h]
    \centering
        \includegraphics[width=1\textwidth]{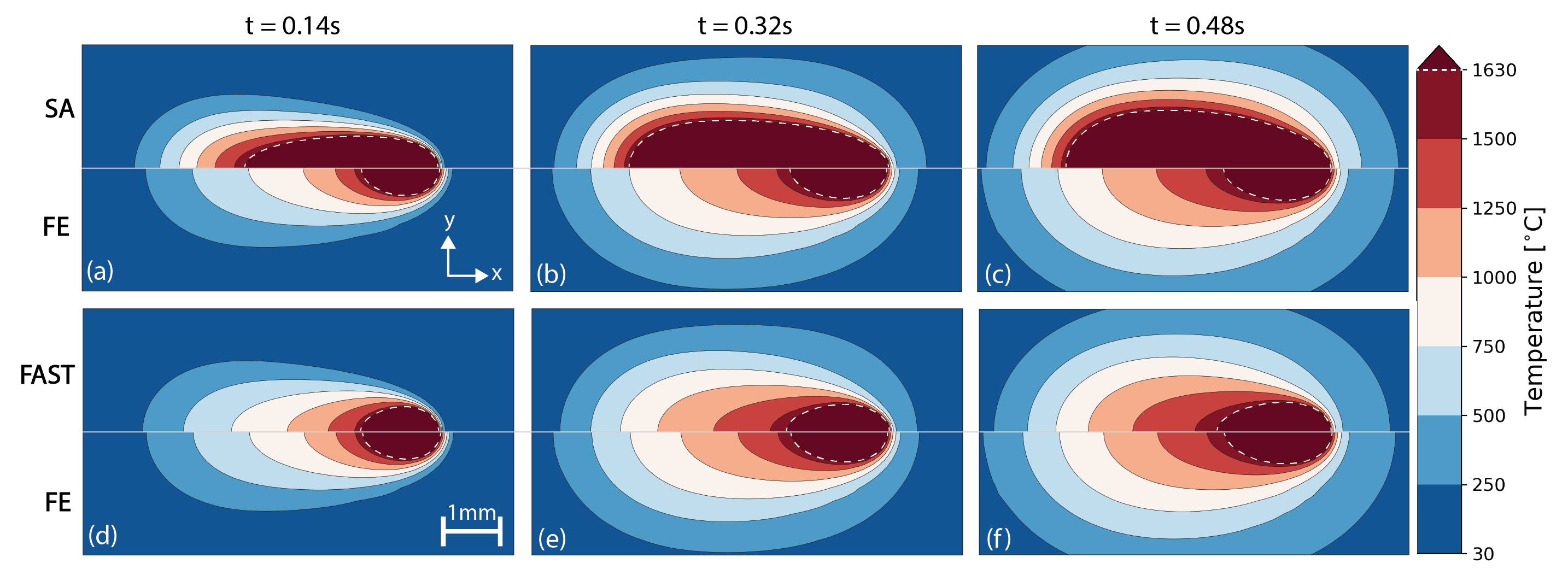}
        \caption{Temperature predicted on the top surface ($z=0$~mm), near the beam position, for reversing melt track pattern at three different times (three columns) for the original SA model (properties evaluated at 35$^\circ$C), the FAST model with positional dependent temperature weighted material properties and the full FE model. Simulation performed for $Q$ = 300 W, $v_B$ = 50 mm/s (6 J/mm) and a background temperature of $T_0$ = 35$^\circ$C. a) and d) correspond to the end of the first track, b) and e) the third track and, c) and f) the end of the last track. The temperatures at the liquidus point, $T_L$, are highlighted with a white dashed line.}
        \label{fig:SA_compare_surf}
    \end{figure}

Figure~\ref{fig:SA_compare_surf} shows a comparison between predicted temperatures and melt track size for the original semi-analytical (SA) model (without correction for radiation and with properties evaluated at $T_0$), for the FE model (radiation and temperature-dependent properties included) and the FAST model. In this case, simulations were performed for $Q$=300~W and 50~mm/s (6~J/mm). Temperature contours on the top surface ($z$=0) are shown for each of the models at the end of the first ($t$=0.14~s), third ($t$=0.32~s) and fifth (last) ($t$=0.47~s) passes.  These correspond to `Track 1', `Track 3' and `Track 5' in Figure \ref{fig:path}.  Comparing, one can see a build up of temperature and a corresponding increase in melt pool size (approximated by the size of the contour evaluated at the liquidus temperature, $T_L$) for all models. It is clear that the original SA model significantly over predicts the temperatures close to the melt pool and, correspondingly the melt pool size when compared to the FE model. It is also notable that the SA predictions worsen as time goes on.  In contrast, the FAST model does a much better job of predicting the near-melt pool temperatures and melt pool size, the prediction fidelity being nearly the same in the first track as in the last. The FAST model performs most poorly in regions where the temperature gradients are largest, particularly just ahead of the beam position at short times where the temperatures ahead of the beam path are lowest.  This corresponds to the results shown in Figure \ref{fig:props_compare} where it was seen that the correction method for material property estimation was poorest in the region just ahead of the beam position. 

\begin{figure}[!h]
    \centering
        \includegraphics[width=0.75\textwidth]{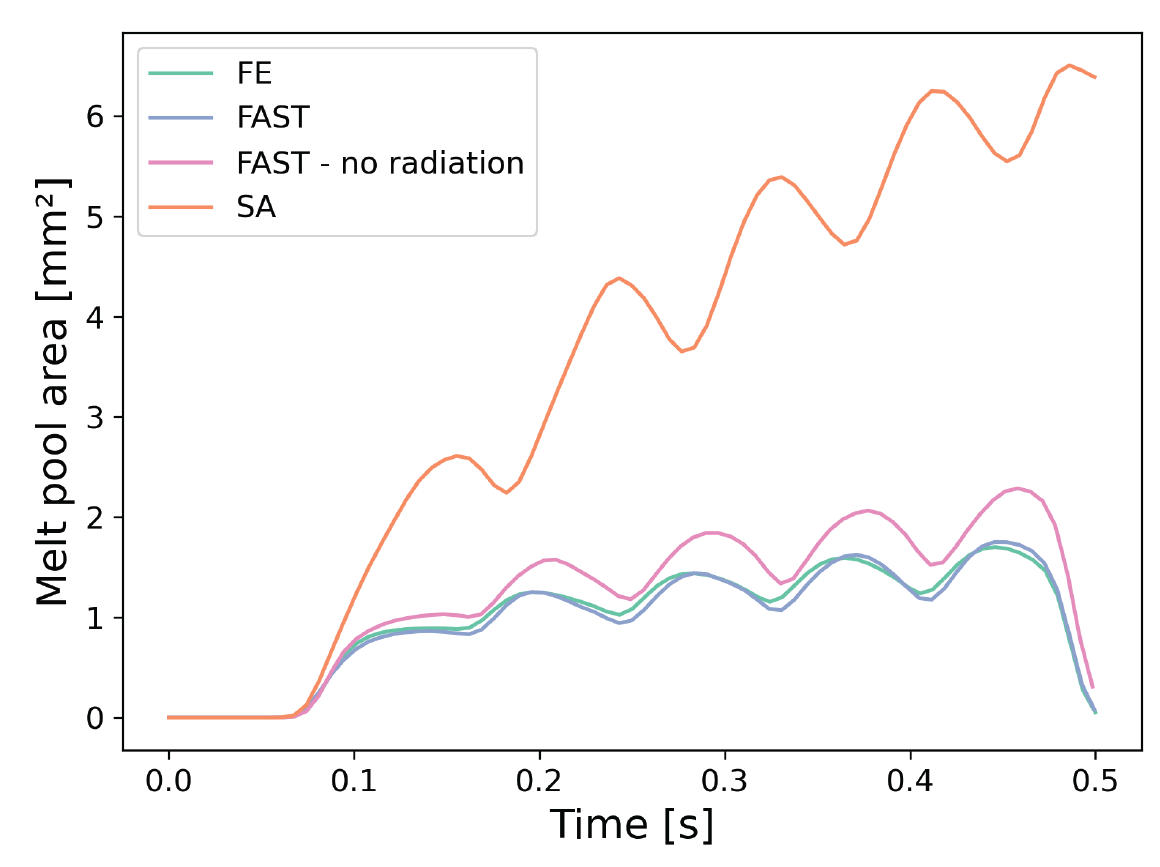}
        \caption{Melt pool area comparison of models for Ti-6Al-4V at $Q$ = 300 W, $v_B$ = 50 mm/s (6 J/mm), and $T_0$ = 35$^{\circ}C$. The melt pool area is defined as the area of the region on the surface inside the contour of the liquidus temperature, $T$ = $T_L$.}
        \label{fig:rad_compare}
    \end{figure}

Looking closer at the change in the melt pool, Figure~\ref{fig:rad_compare} shows the predicted surface area enclosed by the contour $T=T_L$ (as a proxy for the melt pool size) as a function of time for the three models. As one can see, the original SA model diverges from the results of the other two models as heat builds up in the simulation domain.  The role of radiation in particular is highlighted where the FAST model without radiation correction is compared to the full FAST model and the FE simulation.  While all three models predict a steady increase in melt pool size, the FAST model without radiation correction starts to diverge from the FE simulation while the fully corrected FAST model tracks the FE results across the full range of times simulated helping to illustrate the importance, particularly for longer builds, of including the correction for radiative heat loss.

\begin{figure}[!h]
    \centering
        \includegraphics[width=1\textwidth]{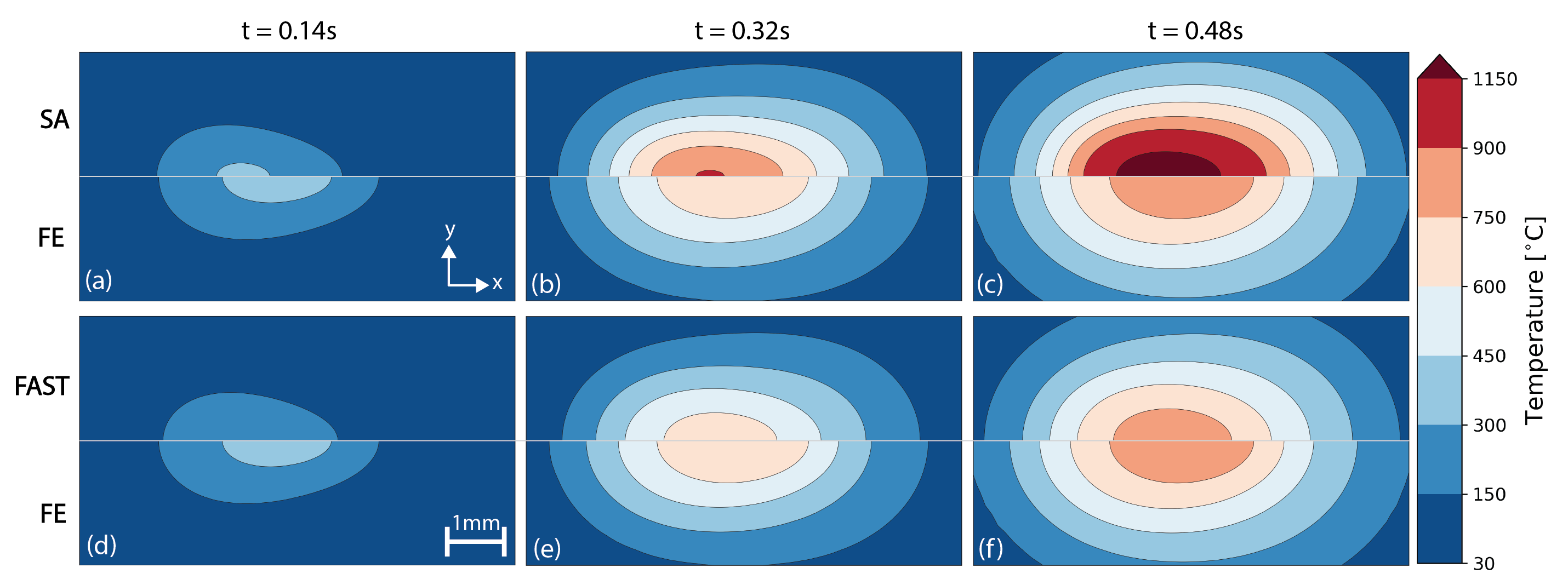}
        \caption{Temperature predicted 1~mm below the top surface ($z=-1$~mm), under the beam position, for reversing melt track pattern at three different times (three columns) for the original SA model (properties evaluated at 35$^\circ$C), the FAST model with position dependent material properties and the full FE model. Simulation performed for $Q$ = 300 W, $v_B$ = 50 mm/s (6 J/mm) and a background temperature of $T_0$ = 35$^\circ$C. a) and d) correspond to the end of the first track, b) and e) the third track and, c) and f) the end of the last track.}
        \label{fig:SA_compare_1mmbelow}
    \end{figure}

The results shown above focus on the temperatures close to the melt pool, on the top surface of the part. As noted above, our aim is to be able to use the FAST model to perform predictions both near-field and far-field with similar levels of accuracy.  To this end, Figure~\ref{fig:SA_compare_1mmbelow} illustrates contour plots for the SA, FAST and FE models on a surface at $z$ = 1~mm below the top surface of the part. These contour plots largely mirror the results shown in Figure~\ref{fig:SA_compare_surf}.

\begin{figure}[!h]
    \centering
        \includegraphics[width=1\textwidth]{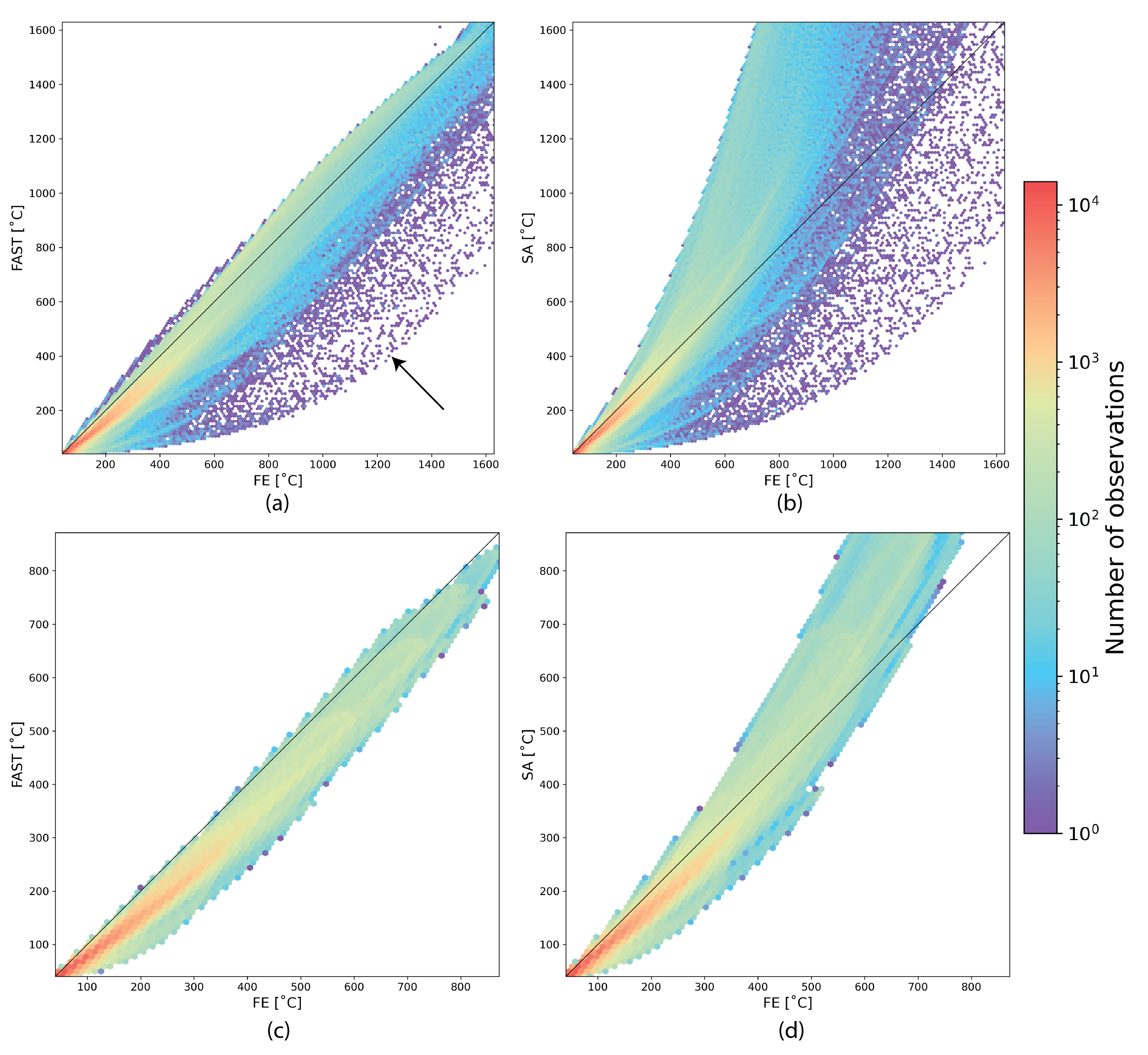}
        \caption{Temperature at all points along surface against one another for a) FAST and FE, b) SA and FE. Similarly, along the plane $z=-1$mm for c) FAST and FE, d) SA and FE. Simulation performed for $Q$ = 300 W, $v_B$ = 50 mm/s (6 J/mm) and a background temperature of $T_0$ = 35$^\circ$C}
        \label{fig:error_compare}
    \end{figure}

To provide a coarser comparison between the three models, over the full time range of the simulations, Figure~\ref{fig:error_compare} shows the temperature at all locations calculated with the FAST/SA models compared to the temperature predicted by the FE simulation at the same locations for 15 different times corresponding to the beginning middle and end of each pass of the beam. Here, only those temperatures at locations where $T \leq T_{L}$  in the FE simulation are considered. In addition, a minimum temperature threshold was used to select only those points whose temperatures were 15\% higher than $T_0$ so that points in the domain unaffected by the heat source do not artificially skew the results. Figure~\ref{fig:error_compare} (a) and (b) show plots for the FAST and SA models for temperature on the surface $z=0$. Here one can see the strong over-prediction of temperatures by the SA model, particularly for locations where the temperatures are predicted to be high. The FAST model shows much better agreement, though with a tendency for under-prediction (negative deviation from line of slope 1). Figure~\ref{fig:error_compare} (c) and (d) show the same plots but now for temperatures predicted at $z = 1$~mm below the top surface. Here, as the temperatures tend to be lower, both models tend to perform better, though again the strong over-prediction of the SA model is apparent.  The FAST model again shows near-linear behaviour aligned along a line of slope 1, with an overall negative deviation.

One feature to highlight in the FAST simulation results in Figure~\ref{fig:error_compare}(a) is the cluster of points that spread out along the line highlighted with the black arrow. These points, having a very large under-prediction of temperatures compared to the FE simulation, arise from points at the leading edge of the melt track. These are the points, noted above, where the temperature gradients are largest and where the corrections for the material properties provide the worst predictions.

\begin{figure}[!h]
    \centering
        \includegraphics[width=1\textwidth]{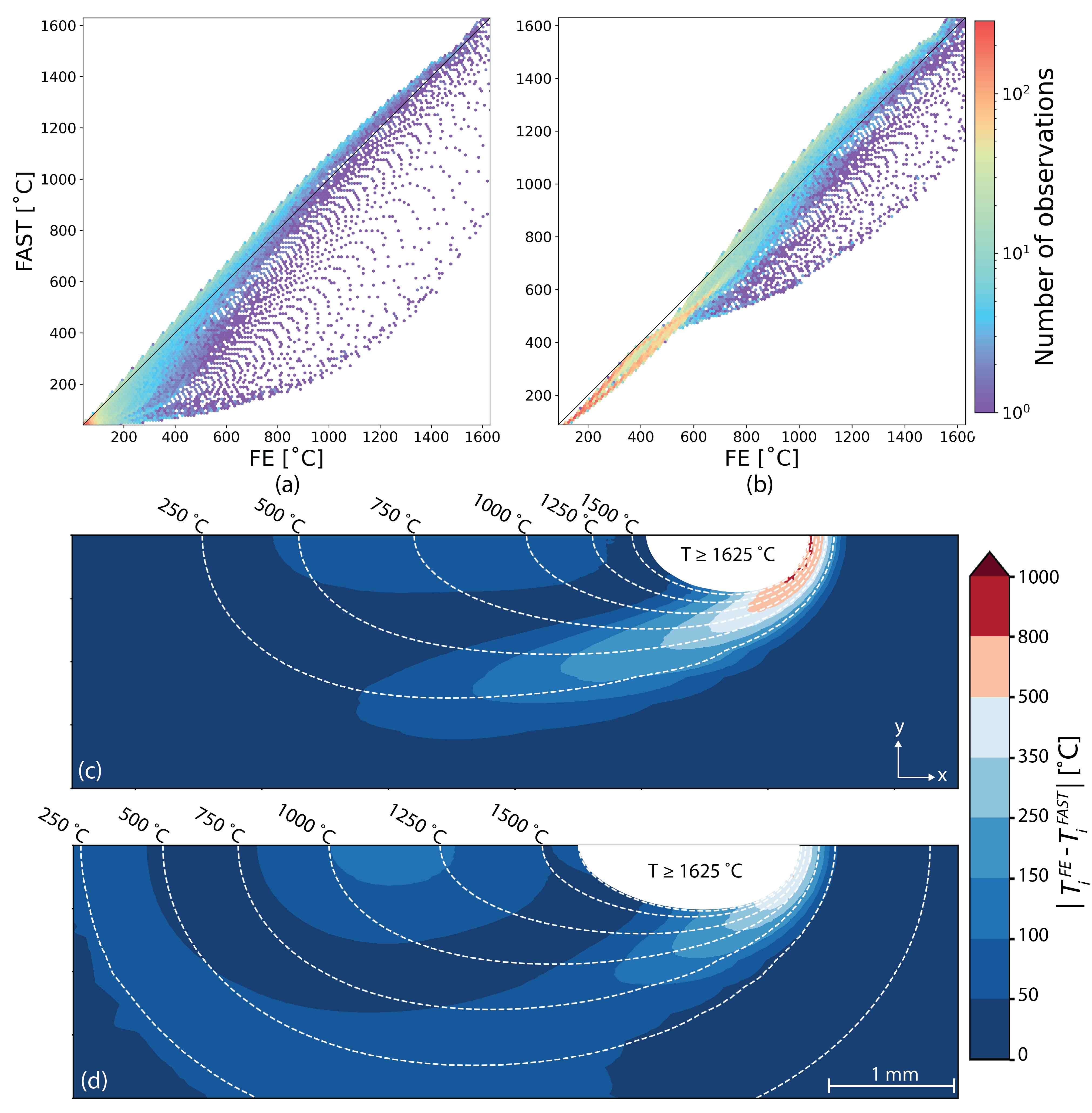}
        \caption{The comparison between the FE and FAST models are shown for a) the end of the first track (track 1) at $t$= 0.16s and b) the end of the last track (track 5) at $t$= 0.48s. The corresponding absolute temperature difference contour plots for c) track 1 and d) track 5 also have the corresponding FE temperature contour lines (white dashed lines) to illustrate the concentration of higher error at the leading edge of the temperature field, specifically for the first track.} 
        \label{fig:error_grad}
    \end{figure}

To better understand the largest deviations, Figure~\ref{fig:error_grad} shows comparisons between FE and FAST model predictions in the same format as in Figure~\ref{fig:error_compare} but now for single times - Figure~\ref{fig:error_grad}(a) t = 0.16 s at the end of the first beam pass and Figure~\ref{fig:error_grad}(b) t = 0.48s at the end of the fifth beam pass. Figure~\ref{fig:error_grad}(c) and (d) show the magnitude of the difference in temperature predicted by FE and FAST models (filled contours) as well as the FE predicted temperatures at the same times as a) and b) in white dashed contour line. The white area where the temperature contours are greater than the melting point are shaded in white to illustrate the area of the melt pool. Here we can see that the largest differences in predicted temperature come at the leading edge of the melt pool where temperature gradients are largest.

The results in Figure~\ref{fig:error_compare} suggest that the FAST model result improves as we move below the surface ($z < 0$). To quantify the error between the models, Equation~\ref{eqn:RMSE} shows the root mean squared error (RMSE) that will be used between predicted temperatures by FE ($T^{FE}_{i}$), and FAST ($T^{FAST}_{i}$) (or SA) at a given point, $i$ = ${x_p,y_p.z_p}$.

    \begin{equation}
    \mathrm{RMSE}=\sqrt{\frac{\sum_{i=1}^N\left(T^{FE}_{i}-T^{FAST}_{i}\right)^2}{N}}
    \label{eqn:RMSE}
    \end{equation}

\begin{figure}[!h]
    \centering
        \includegraphics[width=1\textwidth]{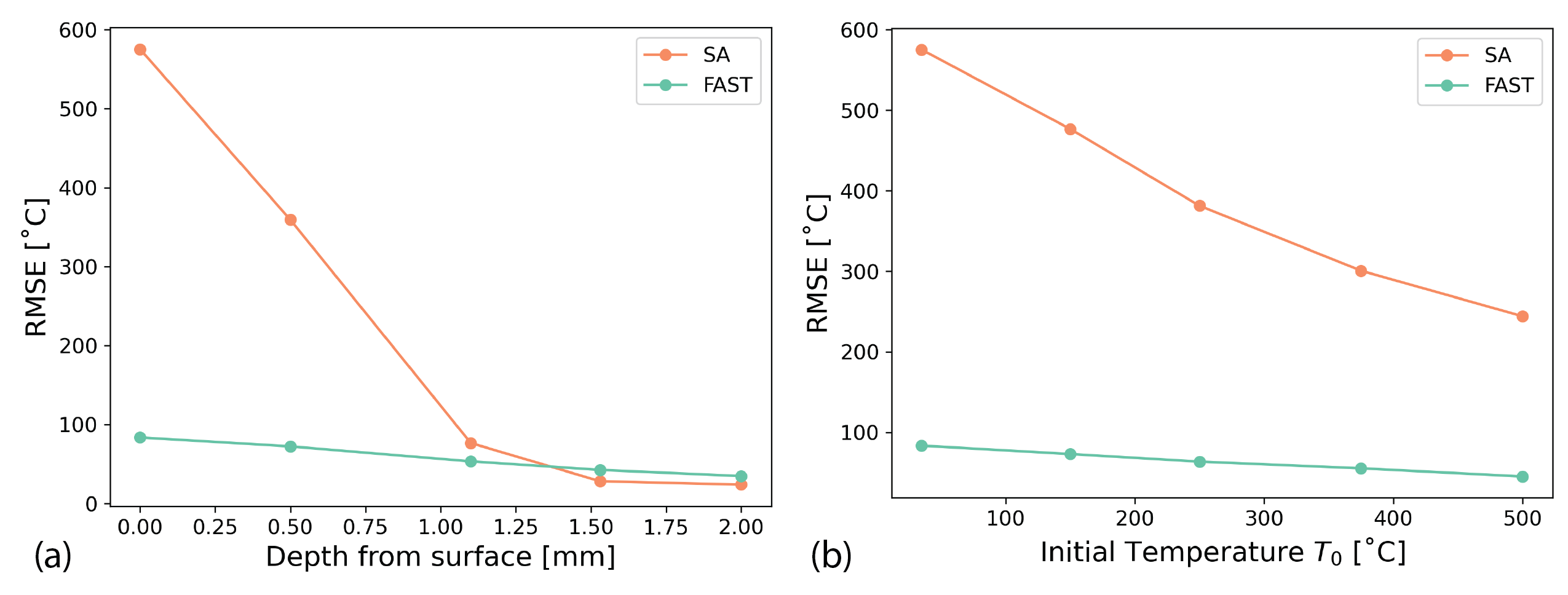}
        \caption{a) RMSE computed between semi-analytical models (SA and FAST) and the FE model on planes at different depths, $z$ and b) RMSE computed on top surface ($z = 0~$mm) as a function of the background initial temperature. In all cases the RMSE has been computed at 15 timesteps, one at the beginning, middle and end of each melt track. Simulation performed for $Q$ = 300 W, $v_B$ = 50 mm/s (6 J/mm) and a background temperature of $T_0$ = 35$^\circ$C}
        \label{fig:RMSE_depthandT0}
    \end{figure}

Figure~\ref{fig:RMSE_depthandT0}(a) illustrates how the root mean square error (RMSE) between the FAST and FE model decreases as depth below surface increases. Similarily, the RMSE for the SA model (relative to the FE model) also decreases, in this case more strongly than the FAST model, leading to the two models giving similar predictions at a distances less than $\sim$1~mm below the surface. This is easily ascribed to the fact that at these distances the material property corrections in the FAST model are much smaller than at $z=0$ since the range of temperatures sampled is much less. This is also seen, though to a smaller extent, by looking at the effect of the initial background temperature $T_0$ (Figure~\ref{fig:RMSE_depthandT0}(b)). Here it can be seen that, as expected, simulations performed at high $T_0$ have much smaller RMSE than those performed at low temperatures.  This, again, can be ascribed to the fact that as $T_0$ increases, the range of variation in material properties is truncated. 

\begin{figure}[!h]
    \centering
        \includegraphics[width=1\textwidth]{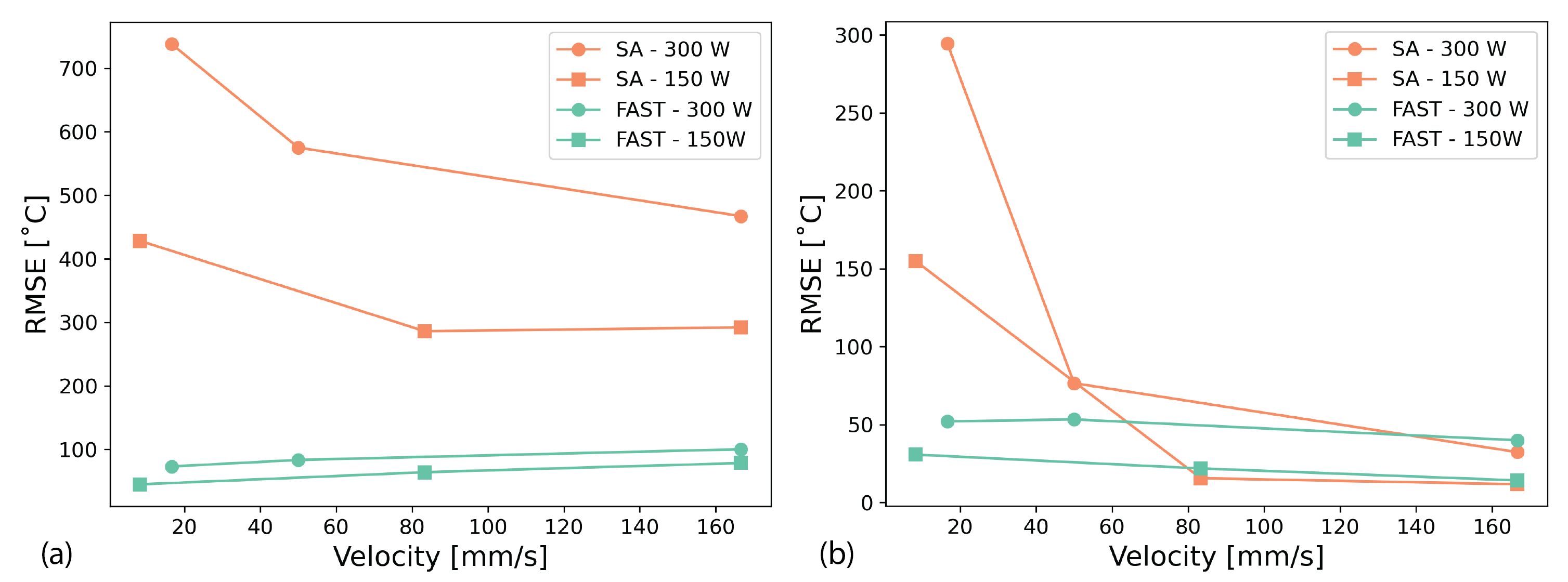}
        \caption{RMSE computed between SA, FAST models and FE model as a function of beam velocity with $Q$ = 150 W and 300 W for a) $z = 0~$mm and b) $z = -1~$mm performed at $T_0$ = 35$^\circ$C.}
        \label{fig:RMSE_VandP}
    \end{figure}

Finally, the above results have been performed for a single combination of beam power and beam velocity of 300 W and 50 mm/s. Figure~\ref{fig:RMSE_VandP} shows how the RMSE evolves with varying beam velocities and powers at $z = 0~$mm and $z = -1$~mm.  For higher beam velocities, the RMSE drops for the SA model as, for the conditions assumed here, the overall temperatures and thermal gradients tend to be lower and so the material property assumptions are more accurate. For all conditions presented here, the RMSE remains nearly the same for the FAST model, as the corrections implemented are able to handle the heat accumulation with similar levels of accuracy for this particular configuration. One does need to be careful with extrapolating these results too far as they do reflect the specific conditions studied here. Under conditions where larger heat accumulation is observed (e.g. smaller domain with zero flux boundary conditions) we would expect the error in the SA model to grow rapidly while the FAST model's error would remain about the same or grow only slightly because of its ability to adapt to the local temperature conditions.

Although the results thus far have been shown using the properties of Ti-6Al-4V depicted in Figure~\ref{fig:FE_compare}(b), the FAST model should be applicable to any metallic system, see e.g. \ref{S:A}.

\subsection{Prediction of Experimental Melt Tracks}

With the FAST model evaluated under a wide range of processing parameters, comparison to experimental melt tracks can now be investigated. To do so, a simple triangle, similar to that modelled by Stump \emph{et al} \cite{stump2019adaptive} has been used. This beam path has a hatch spacing of 0.5~mm and runs 10~mm along the base of the triangle with a 10~mm height in a snake-like trajectory, starting from the bottom left and ending at the apex. The beam trajectory results in a longer time between tracks along the base of the triangle compared to further up the triangle where the time between tracks decreases and the heat is more concentrated and temperatures increase. Two sets of processing parameters were used with a constant beam velocity of 8.33 mm/s and powers of 150W and 200W. Since the absorption efficiency was not measured, an approximate value of 45\% is determined based on fitting to one case and then applied to the other. 

\begin{figure}[!h]
    \centering
        \includegraphics[width=1\textwidth]{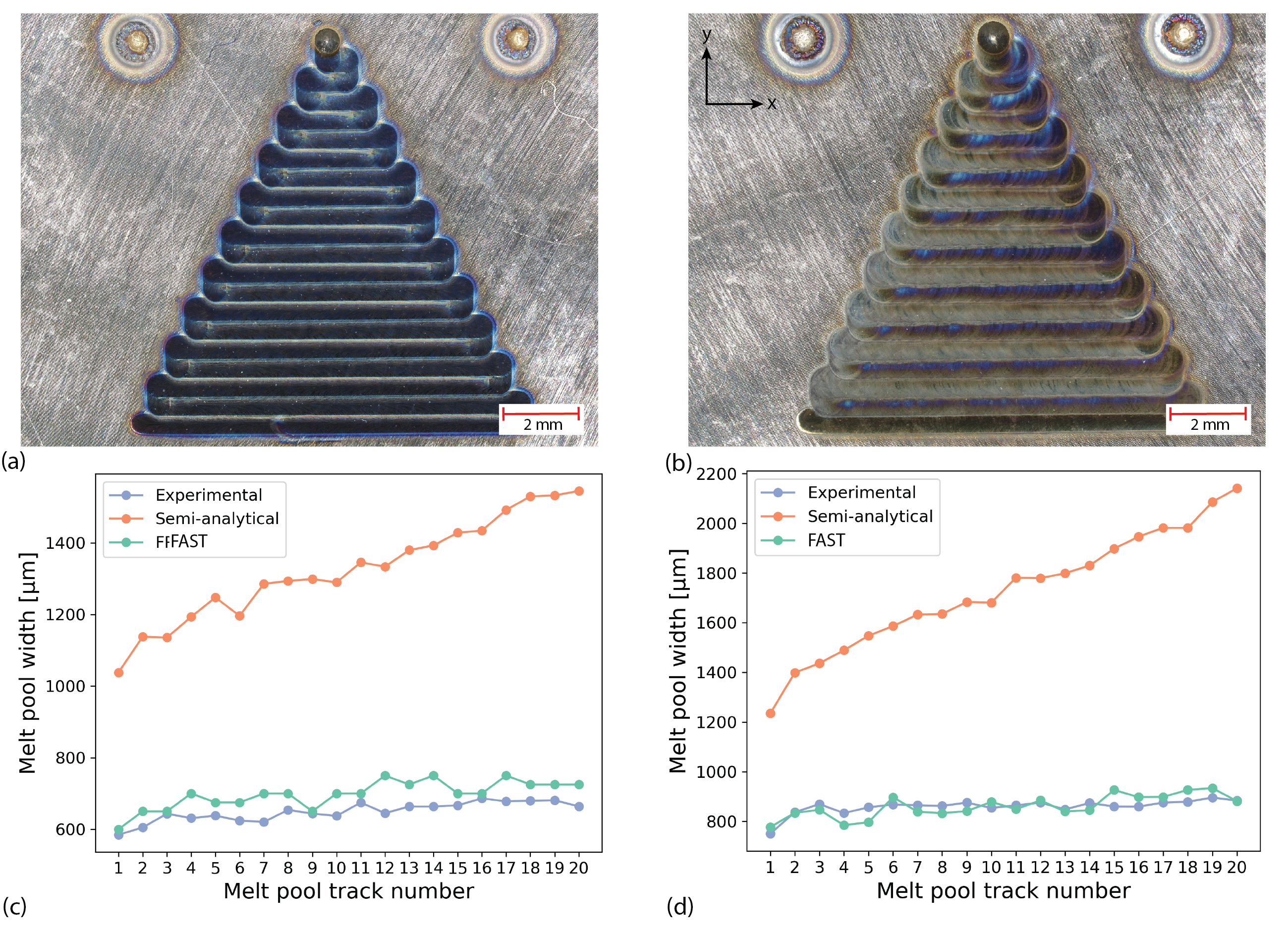}
        \caption{Experimental surface melting for a triangle beam trajectory under a power of (a) 150W and (b) 200W (b) at 8.33 mm/s beam velocity and beam absorption of 45\% while (c) and (d) illustrate the comparison in melt pool widths for the SA and FAST model compared to the experiment for 150 W and 200W, respectively.}
        \label{fig:exp_compare}
    \end{figure}

Surface melting was performed with a stationary fibre laser (wavelength of 1080 nm) and an x-y translation stage to move the solid Ti-6Al-4V substrate. The melt pool widths were measured using a Keyence optical microscope for each individual track throughout the triangle. To differentiate overlapping tracks, the upper bound of the melted profile is measured at the edges of the triangle at each pass. See details in \ref{S:A2}. The melt pool widths for each of the presented models were determined using the FAST model at similar locations as the measurements of each track. The results for each case are shown in Figure~\ref{fig:exp_compare}. 

For both experimental conditions, excellent predictions have been achieved with the FAST model compared to the experimental measurements and a large reduction in error from the original SA formulation is achieved. Moreover, the FAST model is able to capture the trend of heat accumulation much better than the SA model where a significant increase in melt pool width from the first track to the last is observed while in reality there is little change throughout the triangle. This confirms that the FAST model is able to achieve similar errors in experimental measurements as was achieved for the FE model. 

It is also important to highlight the benefits of the semi-analytical models arising from the freedom to choose the number and location of spatial spatial points used. One is not confined to the large domain sizes (and corresponding large number of elements/nodes) required in FE models. Since the spatial domain is decoupled from the temporal domain, one can create a dynamic collection of calculation points, depending on what is most appropriate for the application at hand.

\begin{figure}[!h]
    \centering
        \includegraphics[width=1\textwidth]{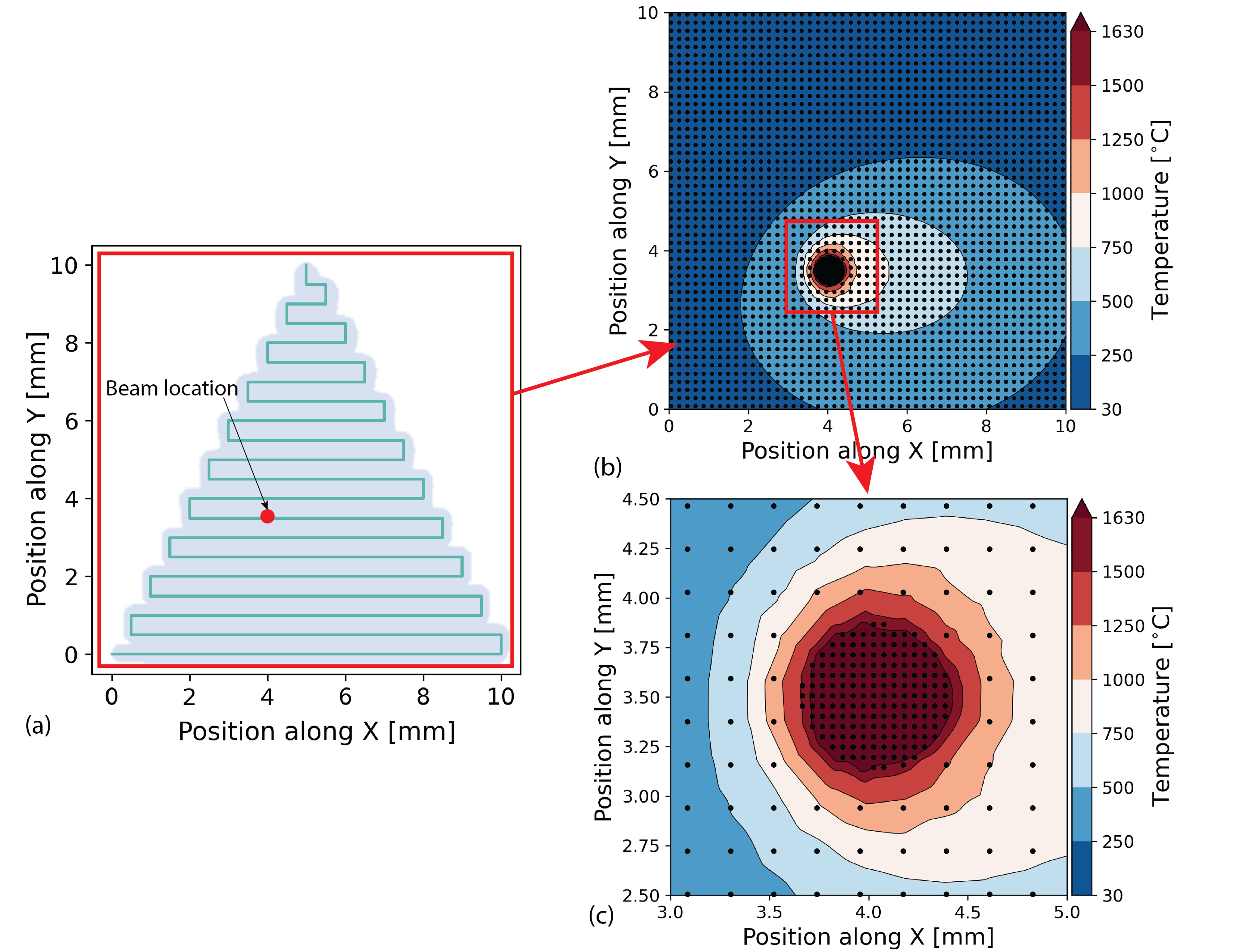}
        \caption{Dynamic mesh capabilities in semi-analytical modelling for the triangle experiment showing for the FAST model: a) the beam trajectory and melted contours, b) a dynamic surface mesh refined around the melt pool, able to resolve long-range and close-range temperatures with a total of 2908 nodes and c) a closer view of the mesh refinement.}
        \label{fig:dyn_mesh}
    \end{figure}

Figure~\ref{fig:dyn_mesh} shows the dynamically refined calculation point capabilities for the SA and FAST models which can be used to reduce computation time while still capturing the areas of interest (e.g. the melt pool width). A dynamic point collection which moves with the heat source and is refined around the areas with high thermal gradients (around the melt pool) is presented in Figure~\ref{fig:dyn_mesh} which highlight this capability. Figure~\ref{fig:dyn_mesh}(b) illustrates the calculation points with a course grid across the entire surface with refinement around the melt pool.

The computational savings compared to FE modelling for the entire 15s build for the SA and FAST models are presented in Table~\ref{tab:comp_time}. The FAST and SA models are compared to the FE model in terms of the how long to run the entire analysis in core-hours using the same domain size as FE of approximately 2.9 million nodes using a static set of points. Although this allows a base-line comparison with the same number of nodes and time steps as the FE model, a more efficient and applicable dynamic mesh, specific to the task at hand should be employed. Here, the dynamic set of calculation points illustrated in Figure~\ref{fig:dyn_mesh} is compared to a similar set of points which only covers the melt pool area and the static grid of points, similar to the FE simulation. Both collection of points are able to resolve the melt pool with the same accuracy but use varying numbers of nodes to do so (2908 and 422 nodes for the dynamic layer points and dynamic melt pool points, respectively) based on the information retained. Since the melt pool widths are measured in the experiment, one can tune the spatial points to output what is pertinent to a targeted evaluation. When running all models with a 16-core Intel(R) Xeon(R) CPU E5-2667 v4 at 3.20~GHz, it is evident that using a similar number of nodes as the FE model in the SA and FAST models results in an order of magnitude decrease in computational savings. Further, when comparing the SA and FAST models, a slight increase in computation time is needed when applying the self-consistent iterative algorithm to account for radiative heat losses. When applying the dynamic spatial set of calculation points compared to achieve the same accuracy to resolve the melt pool, the computation time is reduced by 4 - 5 orders of magnitude compared to the FE model for this case. Here, the FAST model actually runs quicker than the SA model because the smaller melt pool size that results from the radiation correction requires fewer nodes to resolve. 

\begin{table*}[!h]
\centering
\caption{Comparison between the models presented in this study and the relative computation time and speed up compared to FE for various spatial calculation point strategies.}
\label{tab:comp_time}
\begin{tabular}{cccc}
\hline
\addlinespace
\textbf{Model}                           & \textbf{CPU time (core-hours)} & \textbf{Speed up (times)} & \textbf{Computation points} \\ 
\addlinespace
\cline{1-4}
\multicolumn{4}{c}{\textbf{Static calculation points over entire domain}} \\
\cline{1-4}
\addlinespace
FE & 1.2 & - & $2.9 \cdot 10^{6}$ \\ 
SA & $7.0 \cdot 10^{-2}$ & 17 & $2.9 \cdot 10^{6}$ \\ 
FAST & $1.1 \cdot 10^{-1}$ & 11 & $2.9 \cdot 10^{6}$ \\ 
\addlinespace
\cline{1-4}
\multicolumn{4}{c}{\textbf{Calculation points over entire surface with local refinement}} \\
\cline{1-4}
\addlinespace
SA & $1.4 \cdot 10^{-4}$ & 8,571 & 4512 \\ 
FAST & $1.2 \cdot 10^{-4}$ & 10,000 & 2908 \\ 
\addlinespace
\cline{1-4}
\multicolumn{4}{c}{\textbf{Calculation points only over meltpool}} \\
\cline{1-4}
\addlinespace
SA & $8.7 \cdot 10^{-5}$ & 13,793 & 1746 \\ 
FAST & $2.0 \cdot 10^{-5}$ & 60,000 & 422 \\ 
\addlinespace
\hline
\end{tabular}
\end{table*}

In practice, when applied to multi-layer builds, one would likely require the melt pool dimensions (depth and area) as well as possibly a few long range points scattered throughout each layer to capture the majority of the thermal response. Thus, a dynamically refined set of calculation points, similar to the ones presented here would be appropriate to quickly identify problematic areas in the build. It should be noted that the computation time for the SA and FAST models scales approximately linearly with the number of nodes. In the time domain, the scaling is less clear since it depends on the integration scheme, timesteps chosen and the total simulation time. However in the SA model, the computation is done from $t = 0$ to the current time step for each time and thus, with increasing time, the time to compute each subsequent integral also increases. Thus, for applications with long build times for thousands of layers, one may want to investigate further improvements to the time domain rather than just the spatial points.

\section{Summary and Conclusions}

In this study, a methodology to improve the fidelity of a semi-analytical formulation to model the thermal history for a moving heat source is presented.  Previous studies have highlighted the importance of both temperature-dependent  properties and radiation losses to accurately capture to thermal history in AM. Given  that the nature of the SA model neglects these due to its simplifying assumptions, the FAST model was developed to improve the fidelity compared to its equivalent FE  counterpart.
 
Temperature-dependent material properties are implemented using a weighted average of the initial temperature, $T_0$, and the local temperature of the node at the  previous timestep, T$_{local}$. In doing so, the variation in properties spatially is achieved by treating each individual point as its own solution and combining these to accurately represent the thermal history. The effect of radiation was implemented through a self-consistent iterative algorithm using the Boltzmann radiation formulation and corresponding emissivity value for the material. Here, the temperatures at each timestep are calculated and the radiation losses are iteratively scaled down until a converged temperature solution is determined. This heat loss is then removed from  the input power in the following timestep to compensate for the heat that would be lost to radiation.

It was shown through comparisons for a wide range of initial temperatures, powers and beam velocities, that the FAST model maintained a relatively constant  error for temperature predictions throughout the domain compared to its equivalent FE counterpart with radiation and temperature-dependent properties. The validity  of the model was further tested against experimental melt pool width measurements for a triangle beam trajectory. The improvements made to the FAST model had little impact on the computational efficiency and thus maintains the inherent computational efficiency of the SA model with over 4 orders of magnitude speed-up. The work  presented here can now be applied to multi-layer builds and experimental investigation in AM to quickly screen beam trajectories and the resulting characteristics affected by the thermal history.

\section*{Declaration of competing interest}
\noindent The authors declare that they have no known competing financial interests or personal relationships that could have appeared to influence the work reported in this paper.

\section*{Acknowledgments}
\noindent This paper was made possible with funding from the National Sciences and Engineering Research Council of Canada (NSERC) as well as support from William Sparling and Ralf Edinger at Canmora TECH inc.

\appendix
\section{Application to IN718}
\label{S:A}
\begin{figure}[!h]
    \centering
        \includegraphics[width=1\textwidth]{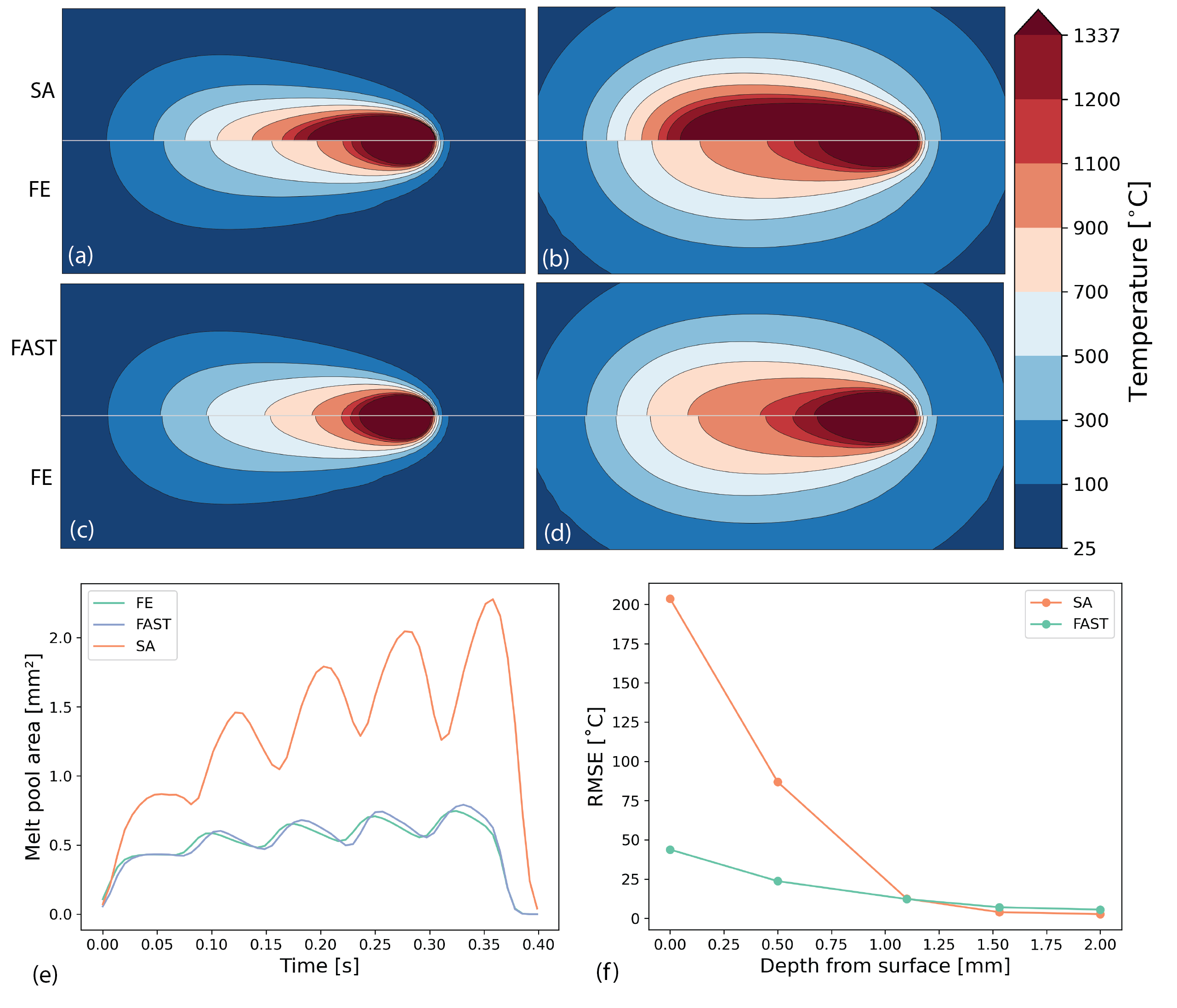}
        \caption{Comparison between temperature contours with IN718 for the SA and FE model for the a) first track and b) last track and comparison between the FAST and FE model for the c) first track and d) the last track. e) shows the melt pool area for the full trajectory for the SA, FAST and FE models is shown in and f) the RMSE as a function of depth for the SA and FAST model against the FE model with $Q$ = 230W, $v_B$ = 3100mm/min and $T_0$ = 35$^{\circ}C$.}
        \label{fig:IN718_compare}
    \end{figure}

To validate the FAST model and its application to other materials, the same beam path for IN718 is modelled and compared with its FE counterpart with both radiation and temperature-dependent properties taken from Ref. \cite{denlinger2016thermal}. The same FE model was used with updated material properties and an emissivity value of 0.5 for IN718 \cite{knapp2019experiments}. The material properties and processing parameters of 230 W, 3100 mm/min and an initial temperature of 35$^{\circ}C$ were chosen based on FE thermal modelling of IN718 for laser PBF study by \cite{denlinger2016thermal}. Figure~\ref{fig:IN718_compare} illustrates contour comparisons for the first and last track as well as the the melt pool area and RMSE through the depth for the beam path. It can be observed that similar to Ti-6Al-4V, the FAST model is applicable to other material systems as well with a significant improvement in the overall thermal history.

\clearpage
\section{Melt pool width measurements}
\label{S:A2}
\begin{figure}[!h]
    \centering
        \includegraphics[width=1\textwidth]{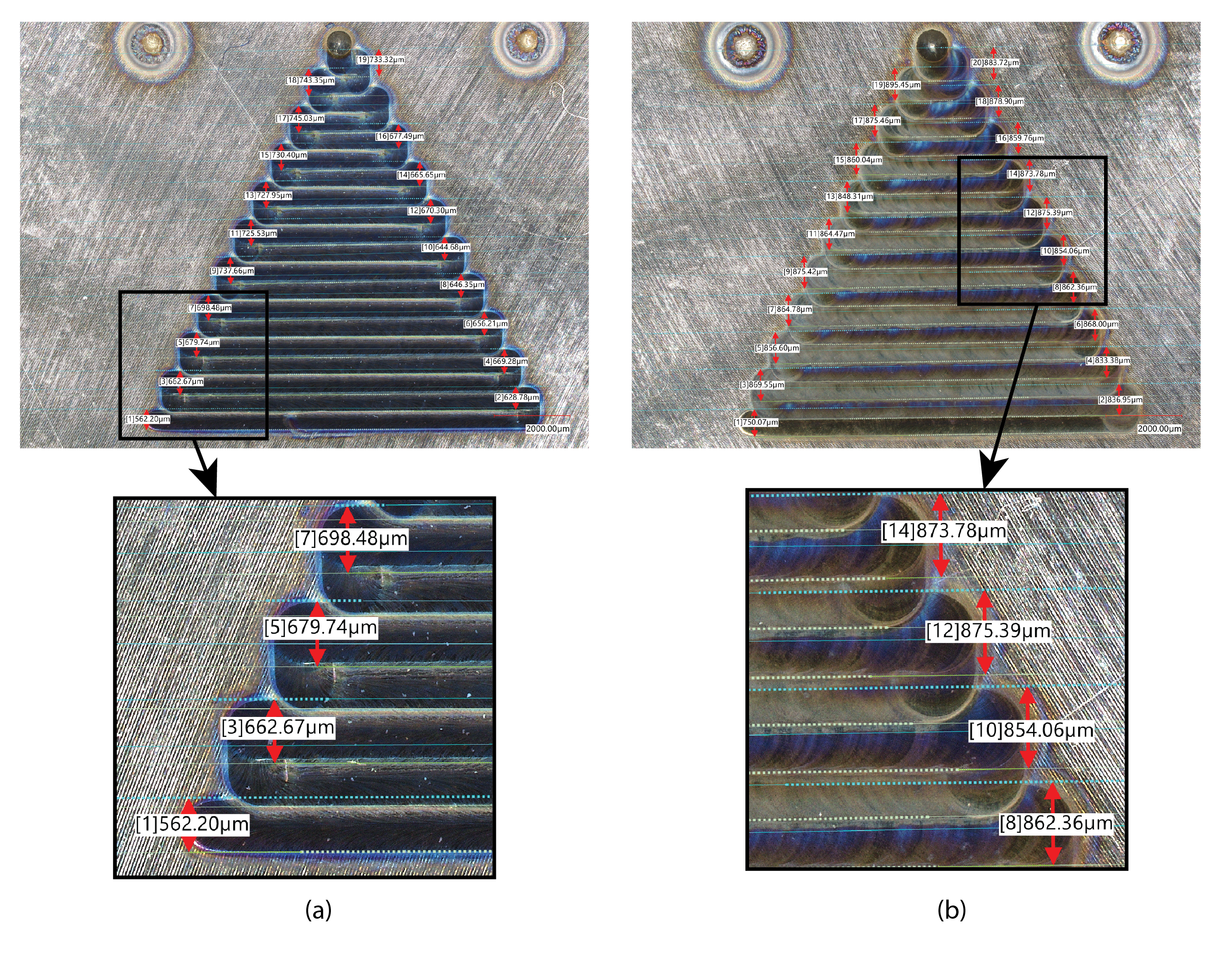}
        \caption{Melt pool width measurements for surface melting of Ti-6Al-4V with a beam velocity of 8.3 mm/s. Each track is recorded just after the corner before the overlapping region from the adjacent melt track for a power of a) 150 W and b) 200W.}
        \label{fig:mpw_measure}
    \end{figure}

Melt pool width measurements were performed with a Keyence VHX-7000 optical microscope for the surface melting on solid Ti-6Al-4V. A triangle with a base of 10mm and a height of 10mm was continuously melted with a hatch spacing of 0.5mm. The beam trajectory starts from the bottom left of the base and end at the apex. A beam power of 150W and 200W with a beam velocity of 8.3 mm/s was used. Although there is overlapping melt profiles, the width of each track was measured at the corner of each melt track, closest to the overlap area form the adjacent melt track. Although this width measurement may not be applicable to the entirety of the track length, the same location was sued to calculated the melt track width in the SA and FAST model for comparison. Figure~\ref{fig:mpw_measure} shows the recorded measurements for each triangle presented in this study.

\clearpage
\clearpage
\renewcommand{\bibname}{References}
\renewcommand{\bibsection}{%
  \section*{\bibname}%
}
\bibliography{citation.bib}

\end{document}